\documentclass[pra,twocolumn,aps,floatfix,showpacs,tightenlines,superscriptaddress,amsmath,amssymb]{revtex4}
\usepackage{amsmath,amssymb,amsbsy,epsfig,color,graphicx,hyperref,bm}

\newcommand{\ket}[1]{|#1\rangle}
\newcommand{\bra}[1]{\langle#1|}

\newcommand{\virg}[1]{``#1''}
\newcommand{\eq}[1]{Eq.~(\ref{#1})}

\def\ff{{\mathcal{J}}}

\begin{document}


\title{Separability and ground state factorization in quantum spin systems}

\author{Salvatore M. Giampaolo}
\affiliation{Dipartimento di Matematica e Informatica,
Universit\`a degli Studi di Salerno, Via Ponte don Melillo,
I-84084 Fisciano (SA), Italy}
\affiliation{CNR-INFM Coherentia, Napoli,
Italy, and INFN Sezione di Napoli, Gruppo collegato di Salerno,
Baronissi (SA), Italy}

\author{Gerardo Adesso}
\affiliation{School of Mathematical Sciences, University of Nottingham,
University Park, Nottingham NG7 2RD, UK}

\author{Fabrizio Illuminati}
\affiliation{Dipartimento di Matematica e Informatica,
Universit\`a degli Studi di Salerno, Via Ponte don Melillo,
I-84084 Fisciano (SA), Italy}
\affiliation{CNR-INFM Coherentia,
Napoli, Italy, and INFN Sezione di Napoli, Gruppo collegato di
Salerno, Baronissi (SA), Italy}
\affiliation{ISI Foundation for
Scientific Interchange, Viale Settimio Severo 65, 00173 Torino,
Italy} \affiliation{Corresponding author. Electronic address:
illuminati@sa.infn.it}

\date{June 4, 2009}

\begin{abstract}
We investigate the existence and the properties of fully separable
(fully factorized) ground states in quantum spin systems.
Exploiting techniques of quantum information and entanglement
theory we extend a recently introduced method and construct a general,
self-contained theory of ground state factorization in frustration free
quantum spin models defined on lattices in any spatial dimension
and for interactions of arbitrary range.
We show that, quite generally, non exactly solvable
translationally invariant models in presence of an external uniform magnetic field can admit exact, fully factorized ground state solutions.
Unentangled ground states occur at finite values of
the Hamiltonian parameters satisfying well defined balancing conditions
between the applied field and the interaction strengths. These conditions are
analytically determined together with the type of magnetic orderings
compatible with factorization and the corresponding values of
the fundamental observables such as energy and magnetization.
The method is applied to a series of examples of increasing complexity,
including translationally-invariant models with short, long, and infinite
ranges of interaction, as well as systems with spatial anisotropies, in low
and higher dimensions. We also illustrate how the general method, besides
yielding a large series of novel exact results for complex models in any
dimension, recovers, as particular cases, the results previously achieved
on simple models in low dimensions exploiting direct methods based on
factorized mean-field ansatz.
\end{abstract}



\maketitle

\section{Introduction} \label{sec.intro}
Quantum information theory is an area of scientific investigation
that has witnessed an enormous progress in the last decade
\cite{NielsChuang}. In the framework of quantum information science,
paradigmatic systems of condensed matter physics such as, e.g.,
spin chains and harmonic lattices, are analyzed from the point of
view of their information content and ability to process and transfer
information. Fundamental concepts of statistical mechanics
and probability, such as the Shannon-von Neumann entropy \cite{Shannon},
play a central role in the quantification of bipartite quantum
entanglement \cite{Horodecki}. On the other hand,
exciting advances in quantum information research
are in a sense paying back the debt: The mathematical and
theoretical toolkit developed for the characterization and
quantification of quantum entanglement has proven useful to tackle
questions and improve our understanding in the investigation of
strongly correlated systems and quantum phase transitions \cite{Amico1}.

Perhaps the most interesting development stemming from the application
of the tools of entanglement theory concerns the study of correlation
scaling in diversely connected systems, with its relations to conformal
field theory and the proof of longstanding conjectures on entropic area
laws in spin and harmonic lattice systems \cite{area}. These
investigations and the associated results are of fundamental interest
because they allow to enhance the control and establish firmer bounds
and limits of applicability on many important tools of simulation and
numerical analysis of complex many-body systems, a fundamental task in
condensed matter, such as the density matrix renormalization group (DMRG)
algorithms \cite{dmrg}, matrix product representations \cite{MPrep},
the entanglement renormalization ansatz \cite{mera}, and weighted graph states methods \cite{weigh}.

The pioneering application of quantum information concepts to
condensed matter was the observation that two-body entanglement,
as quantified by the concurrence, in the ground state of a cooperative
system, exhibits peculiar scaling features approaching a quantum critical
point \cite{FazioNielsen}. These seminal studies helped to clarify that at
quantum phase transitions \cite{Sachdev}, the dramatic change in the ground
state of a many-body system is associated to, or reflected by, a change in
the way that quantum fluctuations are correlated, that is, in the
way entanglement is distributed among the elementary constituents.
Many recent efforts have thus been aimed at understanding the
behavior of different measures of entanglement at quantum phase transitions,
and assessing the corresponding enhancement of properties useful for applications
in quantum technology and quantum engineering \cite{Amico1}. However,
since most entanglement measures can be often rewritten in terms of the
conventional n-point correlation functions, the presence of a divergence in
the former at quantum criticality can be directly traced back to
a divergence in the latter \cite{Campos}, although in some
particular instances entanglement-based studies allowed to discover
novel types of phase transitions \cite{delgadonewwolf}.

Finally, a more technologically oriented product
of the interaction between these two areas of research is the study
of quantum spin chains as natural information carriers and distributors \cite{Bose},
able to realize tasks such
as quantum state transfer \cite{qtransfer} and storage
\cite{giampimemory}, possibly with maximal fidelity
in chains with suitably engineered couplings and dynamics.

Quantum engineering and quantum tech\-no\-lo\-gy are being indeed
developed at a fast pace. Quantum devices are vigorously pursued for
applications ranging from nano-sciences to quantum computation and
entanglement-enhanced metrology \cite{QuantumEngineering}. Despite a
large variety of possible implementations involving different
physical systems, many relevant properties of such devices can be in
fact investigated in a unified setting by appropriate mappings to
quantum spin models \cite{Hartmann,Sorensen}. Control of
separability and entanglement in ground states of quantum spin
models plays an important role in entanglement-enabled quantum
technology applications, and needs to be characterized and
quantified in different physical regimes \cite{Amico1}. From a more
fundamental perspective, the determination of exact solutions
endowed with precisely known properties of separability or
entanglement, can be of great relevance in the study of advanced
models of condensed matter and cooperative systems that are in
general not exactly solvable.

In this paper we exploit quantum information tools to tackle another
important problem in condensed matter physics: the determination and
characterization of exact ground states of cooperative systems
characterized by the property of being in the form of a fully
factorized, tensor product of single-particle pure
states, with no quantum correlations between the
individual constituents. Historically, the occurrence of totally
factorized (unentangled) ground states of quantum many-body systems
was first discovered in the one-dimensional anisotropic Heisenberg
model with nearest-neighbor interactions by Kurmann, Thomas and M\"uller
\cite{Kurmann}, by adopting a direct method in terms of product state
ansatz. This result was later re-derived and extended to two dimensions
using quantum Monte Carlo numerical methods \cite{Roscilde}.
The direct product-state ansatz method of Kurmann, Thomas and M\"uller
has been later extended to anisotropic rings with
interactions of arbitrary range, by exploiting the observation that
factorized ground states break the parity symmetry \cite{Hoeger,Rossignoli}.
Complex quantum systems exhibiting cooperative behaviors, whose ground states are
typically entangled \cite{Typical}, may thus admit, for some non
trivial values of the Hamiltonian parameters, a ground state which
is completely separable. The occurrence of factorization at finite
or even strong values of the couplings is thus an effect of a
delicate balancing between interactions and external fields.

The phenomenon of ground state factorization is particularly
intriguing as it appears to be associated with the presence of an
``entanglement phase transition'' with no classical counterpart
\cite{Amico2}. Furthermore, for the purposes of quantum engineering
applications, that employ distributed entanglement in order to
manipulate and transfer information \cite{Bose}, factorization
points need to be exactly identified and either avoided, in order
to guarantee the reliable implementation of quantum devices, or
properly exploited for the dynamical creation of strongly entangled 
multipartite states of large assemblies of microsystems (graph and
cluster states). Finally, and also quite
importantly, for models not admitting exact general solutions,
achieving knowledge of the exact ground state, even if only for the
restricted nontrivial set of parameters associated to factorization,
would allow (i) to prove the existence of an ordered phase and
characterize it; (ii) to build variational or perturbative
approximation schemes around the exact factorized solution, that may then
be used as test benchmarks for the validity and the precision of
numerical algorithms and simulations. Unfortunately, the direct method
based on the ansatz of a factorized solution is neither sufficiently general
nor mathematically tractable, apart from the notable exceptions of one-dimensional
spin chains with short-range \cite{Kurmann}, infinite-range \cite{Dusuel}, and some types
of long-range exchange interactions \cite{Hoeger,Rossignoli}.

In a recent work \cite{theory} we introduced an analytic method that
allows to determine exactly the existence of factorized ground
states and to characterize their properties in quantum spin models
defined on regular lattices of any size (finite as well as at the thermodynamic limit),
in any spatial dimension, and with spin-spin interactions of arbitrary range.
In correspondence to rigorously established ground state factorizability, the method also
allows to determine novel sets of exact solutions in generally non
exactly solvable models. All the previous particular findings
can be rigorously re-derived and extended
within a unified framework inspired by concepts of quantum
information science.

In the present work we extend the original method introduced
in Ref.~\cite{theory} and we develop a
general self-contained theory of ground state factorization in
(frustration-free) quantum spin systems: With no {\it a priori}
assumption on the nature of the magnetic ordering, we derive a complete
and closed set of conditions that have to be satisfied by the
Hamiltonian interaction parameters in order for the ground state to
be factorized, at precisely determined values of the external magnetic
field, with well defined values of the magnetization and of the energy.
Besides the completely rigorous re-derivation of the few
previously known results on quantum factorization points
\cite{Kurmann,Roscilde,Dusuel,Hoeger,Rossignoli}, our general analytic method
allows to determine novel, exact factorization points and factorized ground
states in, generally non exactly solvable, spin-$1/2$ models as well as in
the corresponding Hamiltonian models of any higher spin. These results hold
true regardless of spatial dimensionality and interaction range, and can
be easily extended even to non-translationally invariant systems. We illustrate
the versatility of the method with a series of different applications,
including models with nearest-neighbor interactions (for which we provide the
full factorization diagram, generalizing the results of Refs.~\cite{Kurmann,Hoeger,Rossignoli}
and of our recent work \cite{theory}), models with long-range interactions
(notably including, among others, the fully connected infinite-range Lipkin-Meshkov-Glick
(LMG) model \cite{Dusuel} and systems with unbalanced next-to-nearest-neighbor
interactions on cubic lattices), and models with spatial anisotropies.

The paper is organized as follows. An excursus on the problem of ground state factorization is presented in Sec.~\ref{sec.History}. In Sec.~\ref{sec.SQUOEXE} we
recall the quantum information toolkit upon which our method is
based, in particular the formalism of single-spin, or single-qubit,
unitary operations  and associated entanglement excitation energies,
previously introduced for the characterization and quantification of
bipartite entanglement from an abstract geometrical perspective
\cite{GiampaoloIlluminati,GiampaoloIlluminatiVerrucchi}. The general method for the rigorous
determination of factorized ground states, that generalizes the
scheme introduced in Ref.~\cite{theory}, is discussed in full detail
in Sec.~\ref{sec.Method}. Relevant applications and examples, including
models defined on lattices of different spatial dimensions and with different
interaction ranges as well as extensions to models with spatial anisotropies
and with dimerized interactions are discussed in Sec.~\ref{sec.short},
Sec.~\ref{sec.long}, Sec.~\ref{sec.lmg}, and Sec.~\ref{sec.anisotropy}.
Concluding remarks and outlooks on future perspectives are presented in Sec.~\ref{sec.Concl}.

\section{Theory of ground state factorization: State of the art}\label{sec.History}

Quantum fluctuations in the ground states of cooperative many-body systems are typically highly correlated, that is, those states contain in the most general instance a strong degree of entanglement distributed among the individual components. This is one of the reasons why their exact characterization is often hard to be accomplished by analytical or even numerical methods. However, there may exist particular solutions corresponding to special values of the Hamiltonian parameters such that the ground state becomes ``classical-like'', in the sense of being a full product state of single-site factors (completely separable, fully factorized state). It is quite surprising that despite the inherent appeal of such exact, special solutions, and the renewed interest stemming from possible applications to quantum technology, very few analytical advances were obtained in the last quarter of a century.

\subsection{The direct method: Product state ansatz}

The first systematic work on ground state factorization in quantum spin models was completed by Kurmann, Thomas, and M\"uller in 1982 \cite{Kurmann}. They focused on the one-dimensional nearest-neighbor anisotropic Heisenberg-like antiferromagnetic spin-$1/2$ model in transverse field, with the assumption that all the spin-spin couplings take nonnegative values. Kurmann {\it et al.} identified a precise value of the external magnetic field, the ``factorizing field'', corresponding to which the system admits a fully factorized ground state. They employed a very direct, ``brute force" method, i.e. formulating a product state ansatz and verifying if and for what values of the Hamiltonian parameters it satisfies the stationary Schr\"odinger problem with lowest energy. Even though this approach appears obvious and extremely easy to pursue, the positive result obtained by Kurmann {\it et al.} in the case of the 1-D Heisenberg model was enabled by some special properties of the case they analyzed. In fact, in order to verify the eigenvalue equation for the product state ansatz, one needs to decompose the total Hamiltonian of the system into a series of pairwise terms, which could be done easily for the model studied by Kurmann {\it et al.} since all the ``pairwise components'' of the factorizing magnetic field had the same weight in this particular case. However, this direct method of determining factorization is in fact significantly less obvious, useful, and comprehensive, when one tries to apply it to more general cases. For instance, considering models with interaction terms of growing spatial range, and/or increasing the spatial dimensionality of the lattice, and/or treating extensions to spatially anisotropic or non-translationally invariant models, results in increasingly hard ``guesses'' concerning the partitioning of the external field into pairwise components and in increasingly nontrivial verification steps. Depending on the structure of the model and on the form of the product state
ansatz, the direct method may either fail to detect factorization points at all, or it may detect only a subset of all physically realizable factorized solutions, or finally it may lead to intractable sets of conditions. These limitations of the direct method become evident, as we will see in the following, even in the simplest 1-D $XYZ$ model with nearest-neighbor interactions analyzed by Kurmann and coworkers. In their original work, due to the insufficient generality of their proposed product ansatz, they failed to identify a wide range of situations that allow the existence of factorized ground states. As we will see when discussing models with competing interactions of different spatial ranges, a further serious fault of the direct method, that becomes incurable in the presence of frustration, is its incorrect assessment of factorized ground states that are in fact excited energy eigenstates of the system.

Following the seminal analysis by Kurmann {\it et al.}, the direct method has been applied to the fully connected infinite-range Lipkin-Meshkov-Glick model by Dusuel and Vidal \cite{Dusuel}, and to the anisotropic Heisenberg chain with the same interactions of arbitrary range along the three spatial directions by Hoeger et al. \cite{Hoeger} and by Rossignoli {\it et al.} \cite{Rossignoli}. We would like to remark that these works, that include the results of Refs.~\cite{Kurmann,Dusuel} as special cases, provide an interesting characterization of factorization points in finite (one-dimensional) translationally invariant lattice spin models with periodic boundary conditions as those points where the parity symmetry is broken.

\subsection{The quantum informatic approach}

A general and rigorous approach to the problem of ground state separability, completely different from the direct method based on the product state ansatz, has been introduced by us in Ref.~\cite{theory} and is fully developed in the present work. This analytic method allows {\it inter alia} to complete the factorization diagram of the anisotropic spin-$1/2$ models without restrictions on the sign of the couplings, and to go significantly beyond the limitations of the previously mentioned studies. Our approach is based on an additional characterization of product states (including product ground states) that employs concepts and tools of quantum information and entanglement theory. The strong point of the quantum informatic approach to ground state factorization lies in its generality and its ensuing independence from the types and ranges of interactions present in the Hamiltonian, the lattice size, and the spatial dimension, which makes it applicable in the most general cases and therefore especially useful in all the situations that are either intractable or incomplete if one resorts to the direct method or to numerical techniques.

In fact, in the original formulation of the quantum informatic method \cite{theory} there was still trace of a residual ``ansatz'' in the initial stage, since the magnetic ordering of the candidate factorized ground state had to be in some sense guessed, and imposed {\it a priori}. Moreover, in the final stage, we were not yet able to derive a completely general set of conditions in order to determine in each situation whether a candidate factorized state that turned out to belong to the spectrum of the Hamiltonian was actually a ground state or belonged to the set of excited states. This had then to be decided case by case in each specific instance.

In the present work we proceed further and develop the complete method to determine and characterize factorized ground states in quantum cooperative systems. We provide the systematics that allows to determine the occurrence of full ground state separability given a generic quantum spin Hamiltonian. The magnetic ordering is derived as a consequence of the theory and not imposed {\it a priori}. A complete set of general conditions is established for the verification of ground state factorization, and the elegance of the method does not hinder practical usefulness in the application to specific models and situations.

We would like to recall that numerical advances on the problem of ground state factorization have been realized by the Monte Carlo study of ground state factorization in two-dimensional anisotropic nearest-neighbor spin models \cite{Roscilde}. Unfortunately, beyond this important result the numerical approach appears to have a very restricted potential for extensions, since the required computational power increases rapidly with the spatial dimension and/or with the range of interactions. We will show that the difficulties inherent to the direct analytic approach and to the numerical method are intrinsically absent in our formalism.

\section{Single qubit unitary operations and entanglement
 excitation energies} \label{sec.SQUOEXE}

In the present section we provide a brief overview of some recent results on the geometrical
interpretation of the entanglement measure known as ``tangle'' (linear entropy) \cite{GiampaoloIlluminati} and the consequences it entails for the relationship between energy and pairwise entanglement \cite{GiampaoloIlluminatiVerrucchi}. Such results embody the premises for our all-analytic characterization of ground state separability in quantum spin systems.

Here and in the following we consider generic systems made of $N$ spin-$1/2$ elementary constituents, or, in the language of quantum information, $N$ qubits. For such a system, we introduce the set of single-qubit unitary operations ({\em SQUOs}) $U_{k}$ \cite{GiampaoloIlluminati}, defined as the unitary transformations that act separately as the identity on any spin of the system except the $k$-th one, on which they act instead as  Hermitian, unitary, and traceless operators:
\begin{equation}
U_{k} \equiv \bigotimes_{i \neq k} \textbf{1}_i \otimes  2 {O}_k \;.
\label{e.Uk}
\end{equation}
In \eq{e.Uk} $\textbf{1}_i$ stands for the identity operator on the $i$-th spin of the system while the generic Hermitian, unitary, and traceless operator ${O}_k$ can be expressed as a linear combination
of the standard spin operators ($S_\alpha^k$) defined on the $k$-th spin:
\begin{equation}
\label{e.Ok} {O}_k=\sin \theta_k \cos\varphi_k S_k^x +\sin\theta_k
\sin\varphi_k S_k^y +\cos\theta_k S_k^z \, .
\end{equation}
In the previous definition the parameters $\theta_k$ and $\varphi_k$ take values respectively in the ranges $(-\pi/2, \pi/2]$ and $[0,2\pi)$.
Physically, $O_k$ corresponds to a rotation in the spin space, and the traceless condition imposes the rotation operation to be orthogonal to the identity, so basically a combination of spin flip and phase flip.

A SQUO transforms every pure state $\ket{\Psi} \in {\mathbb{C}}^{2N}$ into a new state $\ket{\tilde{\Psi}}=U_{k}\ket{\Psi}$, that in general differs from
$\ket{\Psi}$. We may quantify the action of the SQUO in terms of the trace distance between the original and the transformed state, defined as $D(U_k;\ket{\Psi})= D(\theta_k ;\varphi_k
;\ket{\Psi}){\equiv} \sqrt{1{-} |\bra{\Psi}U_k\ket{\Psi}|^{2}}$. The distance $D$ varies in the interval $[0,1]$ and vanishes if and only if the two states coincide, meaning that the considered SQUO leaves the original state unchanged. For a given initial state and for any site $k$, one can determine the {\em Extremal-SQUO} (E-SQUO) that minimizes such a distance. One can prove that the E-SQUO is uniquely determined by the following conditions on the angular variables \cite{GiampaoloIlluminati}:
\begin{eqnarray}
\bar{\varphi}_{k} & = & \arctan{\left( \frac{\langle S^y_k \rangle
}{\langle S^x_k \rangle } \right)}
\; , \nonumber \\
& & \nonumber \\
\bar{\theta}_{k} & = & \arctan{\left( \frac{\langle S^x_k \rangle
\cos{\bar{\varphi}} + \langle S^y_k \rangle
\sin{\bar{\varphi}}}{\langle S^z_k \rangle} \right)} \; ,
\label{e.minimumpoint}
\end{eqnarray}
where $\langle S^\alpha_k \rangle$ denotes the expectation value of the spin operator $S^\alpha_k$ on the unperturbed state $\ket{\Psi}$ ($\langle S^\alpha_k \rangle=\bra{\Psi}S^\alpha_k\ket{\Psi}$). The
E-SQUO plays a crucial role in our analysis: For any pure state defined on a general system of interacting qubits one can prove \cite{GiampaoloIlluminati} that the square of the distance associated to the action of the E-SQUO coincides with the linear entropy $S_L(\rho_{k}) = 1 - Tr[\rho_k^2]$ of the $k$-th qubit (where $\rho_k$ denotes the reduced density matrix of qubit $k$). This quantity, also known as the {\em tangle} $\tau$ in the literature, is an entanglement monotone for qubit systems and satisfies important monogamy constraints \cite{Coffman}. Being a monotonic function of the von Neumann entropy of entanglement \cite{Bennett}, it quantifies the bipartite block entanglement present in the state $\ket{\Psi}$ between spin $k$ and the remaining $N-1$ constituents, and is thus a proper measure of single-site entanglement \cite{Larsson}. One has the following relation \cite{GiampaoloIlluminati}:
\begin{widetext}
\begin{equation}\label{e.tangle}
D^{2}(\bar{U}_k;\ket{\Psi})= \tau_{k|\forall j \neq k}(\ket{\Psi}) \equiv
4\mathrm{Det}\rho_k = 1 -4
\left[ \langle S^x_k \rangle^2 + \langle S^y_k \rangle^2 + \langle
S^z_k \rangle^2 \right] \; .
 \end{equation}
\end{widetext}
This geometric-operational characterization of the tangle allows to develop a quantum informatic approach to the problem of characterizing the correlation properties of the ground state in many-body quantum systems \cite{GiampaoloIlluminatiVerrucchi}. Let us consider a collection of spin-$1/2$ systems defined on a regular lattice, with Hamiltonian $H$, and let $\ket{\Psi}$ denote from now on the ground state of $H$. If the system is in the ground state, the application of a SQUO on a generic spin of the lattice is necessarily associated to an energy transfer that one must provide to the system in order to realize the selected SQUO. In other words, a SQUO perturbs the system, and this perturbation results in an increase of the average energy (obviously, in general the transformed state needs not to be an eigenstate of the Hamiltonian). Such energy deviation from the ground-state value can be quantitatively defined as
\begin{equation}
\Delta E(U_k) \equiv\Delta E(\theta_{k},\varphi_{k})=
\bra{\Psi}U^\dagger_k{H}U_k\ket{\Psi} - \bra{\Psi} {H}\ket{\Psi} . \label{e.EEs}
\end{equation}
Because $\ket{\Psi}$ is the ground state of $H$, $\Delta E(U_k)$ is a non negative defined quantity. Considering the special relation between the E-SQUO and the tangle we may expect that the associated amount
of energy difference $\Delta E(\bar{U}_k)$, appropriately named `Entanglement eXcitation Energy' (EXE), will have as well a strong, direct connection with the tangle and in general with the single-site entanglement in the ground state. In fact, starting from the above definition of \eq{e.EEs} and taking into account the property of the E-SQUO of leaving unchanged a fully disentangled state, it has been established in full generality \cite{GiampaoloIlluminatiVerrucchi} that if the system is invariant under spatial translation and the Hamiltonian of the system does not commute with any possible SQUO, then the vanishing of the EXE is a {\em necessary and sufficient} condition to admit a fully factorized ground state $\ket{\Psi}$ \cite{GiampaoloIlluminatiVerrucchi}. Hence, if the system that we want to analyze satisfies all the former hypotheses, the vanishing of the EXE provides an additional condition, besides the vanishing of the tangle $\tau$, that one needs to include in establishing a general analytic approach to the problem of ground state factorization.


\section{Theory of ground state factorization: General analytic method}\label{sec.Method}

\subsection{Preliminaries}

The general method that was first partially developed in Ref.~\cite{theory} for the determination of existence, location, and exact form of factorized ground states rests on the following two main observations:

\begin{description}
\item[\rm (i)] The ground state $\ket{\Psi}$ of a spin-$1/2$ Hamiltonian $H$
is factorized if and only if the single-site entanglement (tangle) $\tau$ vanishes for all
spins $k$ in the lattice; \item[\rm (ii)] Provided that $[H,U_k]\neq
0$ for every possible SQUO $U_k$ of the form \eq{e.Uk}, the ground
state $\ket{\Psi}$ of a spin-$1/2$ Hamiltonian $H$ is factorized if
and only if the EXE $\Delta E({\bar U}_k)=\bra{\Psi} {\bar U}_k H
{\bar U}_k \ket{\Psi} - \bra{\Psi} H \ket{\Psi}$ vanishes $\forall
k$.
\end{description}

In Ref.~\cite{theory}, the strategy to the determination and
understanding of factorization for a given Hamiltonian bore on the
above two facts and on the requirement to assume as working point
a phase endowed with some kind of magnetic order. By imposing
the vanishing of the EXE and of the tangle, we showed how to
determine uniquely the form of the completely disentangled state
compatible with the assumed magnetic order, candidate to be the
ground state of the system at a particular value of the external
magnetic field (labeled as ``factorizing field''). The final steps of
the procedure concern the determination of exact analytic
conditions for the candidate factorized state to be an eigenstate of
the Hamiltonian with lowest energy. In Ref.~\cite{theory}, we derived
a closed set of eigenstate conditions, but establishing whether a
factorized eigenstate is actually a ground state had to be left to a
case by case analysis.

In the present paper we generalize the method, extend its scope, and
apply it to different classes of quantum spin models. Firstly, we
will show that there is no need for {\it a priori} assumptions on the
magnetic order. In fact, we will prove that the existence of a factorized
ground state is a necessary and sufficient condition for the existence
of quantum phase transitions and ordered phases in translationally-invariant
quantum spin models with exchange interactions and in the presence of
external fields. Moreover, the actual structure of the factorized ground
state determines automatically the kind of ordered phase.
Further, we derive a complete set of rigorous conditions for
the candidate disentangled state to be an eigenstate and
a ground state. In particular, for systems with no frustration
we prove that if the factorized state is an eigenstate,
then it is always a ground state of the system.

\subsection{Quantum spin models on regular lattices}

We illustrate the method in detail by
considering its application to the general, translationally
invariant, exchange Hamiltonian $H$ for spin-$1/2$ systems on a
$D$-dimensional regular lattices, with spin-spin interactions of
arbitrary range and arbitrary anisotropic couplings:
\begin{equation}\label{e.Hamiltonian}
H=\frac{1}{2} \sum_{\underline{i},\underline{l}} J_x^r
S_{\underline{i}}^x S_{\underline{l}}^x+ J_y^r S_{\underline{i}}^y
S_{\underline{l}}^y +J_z^r S_{\underline{i}}^z S_{\underline{l}}^z-
h\sum_{\underline{i}} S_{\underline{i}}^z \, .
\end{equation}
Here $\underline{i}$ (and similarly $\underline{l}$) is a
$D$-dimensional index vector identifying a site in the lattice,
$S_{\underline{i}}^\alpha$ $(\alpha=x,y,z)$ stands for the
spin-$1/2$ operator on site $\underline{i}$, $h$ is external field
directed along the $z$ direction, $r=|\underline{i}-\underline{l}|$
is the distance between two lattice sites, and $J_\alpha^r$ is the
spin-spin coupling along the $\alpha$ direction. Translational
invariance is ensured by the fact that all the couplings depend
only on the distance $r$ between the spins. This type of
Hamiltonian encompasses a large variety of models and spans over
several universality classes including, among others, the Ising, XY,
Heisenberg, and XYZ symmetry classes. Besides their importance in
quantum statistical mechanics and in the theory of quantum critical
phenomena, these models play an important role in the study of various
schemes of quantum information and quantum communication tasks
with quantum many-body systems \cite{Bose}.

It is a rather straightforward exercise to verify that the Hamiltonian
\eq{e.Hamiltonian} never commutes with the SQUOs
$U_{\underline{k}} \; \forall \underline{k}$. Translational invariance
and non-commutativity of the Hamiltonian with the SQUOs
guarantee that the vanishing of the EXE is a necessary and
sufficient condition for the full separability of the ground
state [statement (ii) above]. Since the external field $h$
is uniform on the lattice, it forces the existence of a
site-independent, nonvanishing magnetization $M_z$ along the $z$ axis:
$\langle S_{\underline{k}}^{z} \rangle \equiv M_z \,\forall \underline{k}$.
On the other hand, at a factorization point, statement (i) above imposes
the vanishing of the single-site entanglement on all sites of the lattice.
Therefore, if a factorization point exists, the magnetizations along
the directions orthogonal to the external field must assume the following
form:
\begin{eqnarray}
\label{e.magnxy} \langle S^x_{\underline{k}}\rangle & \equiv
M_{\underline{k}}^x& = M_\bot \cos \varphi_{\underline{k}}, \nonumber \\
\langle S^y_{\underline{k}} \rangle & \equiv M_{\underline{k}}^y& =
M_\bot \sin \varphi_{\underline{k}},
\end{eqnarray}
where $M_\bot= \sqrt{\frac{1}{4} - M_z^2}$ is the modulus of
the projection of the magnetization on the $xy$ plane and the local
orientations $\varphi_{\underline{k}}$, and hence the type of magnetic
ordering in the $xy$ plane, remain undetermined. Since $M_z$ is site-independent,
then $M_\bot$ is site-independent as well.
Taking into account \eq{e.minimumpoint} of
Sec. \ref{sec.SQUOEXE} one sees that, at least at the
factorization point, $\bar{\theta}_{\underline{k}}$, i.e. one
of the two angles that fix the orientation of the E-SQUO, does not
depend on the site index. Therefore $\bar{\theta}_{\underline{k}}\equiv\theta$
$\forall \underline{k}$.

Collecting the former results and considering that in
the presence of a factorized ground state, if it exists, all the
correlation functions are products of single-site
expectations values, we obtain that the condition on the vanishing
of the EXE associated to the generic site $\underline{k}$ reads:
\begin{widetext}
\begin{eqnarray}\label{e.EXE}
\frac{\Delta E(\bar{U}_{\underline{k}})}{2}&=&0 \nonumber \\&=& - \sin
\theta\left( \sin \theta M_z  - \cos \theta M_ \bot \right) \left(
M_z\sum_{\underline{i}} J^z_{\underline{k},\underline{i}} - h_f
\right)+ \nonumber \\
& & +\cos \theta M_ \bot \left( \sin \theta M_z - \cos \theta M_
\bot \right) \left( {\cos \varphi _{\underline{k}}
\sum_{\underline{i}} {J^x_{\underline{k},\underline{i}} \cos \varphi
_{\underline{i}} } + \sin \varphi _{\underline{k}}
\sum_{\underline{i}} {J^y_{\underline{k},\underline{i}} \sin \varphi
_{\underline{i}} } } \right) \, .
\end{eqnarray}
\end{widetext}
In the former equation $h_f$ is the {\em factorizing field}, i.e.
the value, if it exists, of the external field for which ground state
factorization occurs.

Since we are looking for a fully factorized ground state,
\eq{e.EXE} must be satisfied simultaneously  for all
$\underline{k}$ and, henceforth, \eq{e.EXE} must be
site-independent. This fact in turn implies that the following
relation must be satisfied by every angle $\varphi_{\underline{k}}$:
\begin{equation}\label{e.invariance}
\cos \varphi _{\underline{k}} \sum\limits_{\underline{j}} {J_x^r
\cos \varphi _{\underline{j}} } + \sin \varphi _{\underline{k}}
\sum\limits_{\underline{j}} {J_y^r \sin \varphi _{\underline{j}} } = K \, .
\end{equation}
\eq{e.EXE} must hold independently of the site on which the E-SQUO acts
and the value of the constant $K$ must be identified by determining the
expression of the energy density, i.e. the energy per site, associated to the
factorized ground state. Given the total ground-state energy $E$ and the total
number of sites $N$, the latter quantity can be written as
\begin{eqnarray}
\label{e.energy}
\frac{E}{N} &=& \frac{1}{2}M_z^2 \sum\limits_r^{} {Z_r J_z^r }  - h_f M_z + \nonumber \\
& + & \frac{1}{2}M_ \bot ^2 \sum\limits_{\underline{j}}^{} { {J_x^r
\cos {\varphi _{\underline{k}} } \cos {\varphi _{\underline{j}} } }
} { + J_y^r \sin {\varphi _{\underline{k}} } \sin
{\varphi_{\underline{j}} } } \nonumber \\
&\equiv& \frac{1}{2}M_z^2 \sum\limits_r^{} {Z_r J_z^r }  -
h_f M_z + \frac{1}{2}M_ \bot ^2 K\,,
\end{eqnarray}
where $Z_r$ is the coordination number, that is the number of sites
at distance $r$ from a given site. In order to determine the
magnetic order in the candidate factorized ground state,
one can now exploit the fact that, by definition, it
must minimize the energy of the system. Therefore
such state will be simply characterized by the magnetic order that
minimizes the per-site contribution $K$, depending on the
Hamiltonian parameters and the geometric structure of the
lattice. Thus the general conditions for full separability of
the ground state lead to the uncovering of the magnetic order
in the $xy$ plane. In the original formulation of the quantum-informatic
method \cite{theory} one had to assume from the beginning the type
of ordered phase, and this resulted in constraints on the magnetization
and limitations in the range of the coupling coefficients.

\subsection{Magnetic orders and factorization}

The outcome of the minimization of the energy density depends on the
presence or the absence of frustration in the system,
either due to the lattice geometry, or to the structure of the
Hamiltonian, or to both causes. For Hamiltonians belonging to the
class of \eq{e.Hamiltonian}, frustration can be due both to the geometry
of the lattice, e.g. for systems with nearest-neighbor antiferromagnetic
interactions defined on a lattice that admits closed loops with an odd
number of spins, and to the competition of interactions of different
spatial orders, e.g. in systems defined on linear chains with nearest-neighbor
ferromagnetic and next-nearest-neighbor antiferromagnetic interactions.
Frustration effects in systems with interactions of different spatial ranges
arise whenever it happens that the minimization of the energy associated to
interactions over a certain spatial scale precludes the possibility of
minimizing the energy associated to interactions over a different spatial scale.
In the absence of frustration, i.e. when the lattice geometry and the interactions
between the spins are such to allow all energy minimizations simultaneously, it is
straightforward to prove that every possible factorized ground state must be
characterized by one of only four magnetic orders out of the many possible ones.
The complete proof of this statement is reported in Appendix
\ref{appendixfourcases}. This restricted set includes ferromagnetic ordering
along the $x$ direction ($\varphi_{\underline{k}}=0\, \forall \underline{k}$)
or along the $y$ direction ($\varphi_{\underline{k}}=\pi/2 \, \forall
\underline{k}$), and antiferromagnetic ordering along the $x$ direction
($\varphi_{\underline{k}}=|\underline{k}|\pi \, \forall
\underline{k}$) or along the $y$ direction ($\varphi_{\underline{k}}=(\pi/2)+
|\underline{k}|\pi \, \forall \underline{k}$). By requiring the
minimization of the energy per site, \eq{e.energy}, the type of
order effectively present in the ground state can be determined
by comparing the values of the different ``net interactions'' along
the $x$ and $y$ directions, where the net interactions are defined as follows:
\begin{eqnarray}
\label{e.netinteractions}
\ff^A_{\alpha} &=& \sum_{r=1}^{\infty} (-1)^r Z_r J^r_{\alpha}\,; \nonumber \\
\ff^F_{\alpha} &=& \sum_{r=1}^{\infty}  Z_r J^r_{\alpha}\, ,
\end{eqnarray}
with $\alpha=x,y$.
The net interaction along the $z$ direction is defined equivalently as
\begin{equation}
\label{e.netinteractionz}
\ff^{A,F}_{z} = \sum_{r=1}^{\infty}  Z_r J^r_{z}.
\end{equation}
As a function of the net interactions, the magnetic orders in the
system are determined by the values of
$\mu=\min\{\mathcal{J}^F_x,\mathcal{J}^A_x,\mathcal{J}^F_y,\mathcal{J}^A_y\}$:
\begin{equation}
\mu =\left\{ \begin{array}{ccl}
  \mathcal{J}^F_x & \Rightarrow & \hbox{Ferrom. order along $x$ ;}  \\
  \mathcal{J}^F_y & \Rightarrow & \hbox{Ferrom. order along $y$  ;} \\
  \mathcal{J}^A_x & \Rightarrow & \hbox{Antiferrom. order along $x$  ;} \\
  \mathcal{J}^A_y & \Rightarrow & \hbox{Antiferrom. order along $y$ .}
  \end{array}
  \right.
  \label{fourcases}
\end{equation}
It is interesting to observe that, in terms of the net interactions, one
can immediately establish the presence or absence of frustration
in the system: Any frustration arising from the presence of competing
interactions would imply that $\mu \neq -\sum_r Z_r |J_\alpha^r|$,
with $\alpha=x,y$ depending on which of the two axes is characterized
by a non vanishing value of the magnetization. We remark
that here and in the following we are not considering the particular
situation in which a {\em saturation} occurs rather than a true ground
state factorization: This instance will be discussed separately in
SubSec.~\ref{s.sec.saturation}.

From now on, without loss of generality, we will specialize our analysis
to the the case of an antiferromagnetic ordering along the $x$ direction.
With trivial modifications, all the steps and results that will be obtained
in the following hold as well in the remaining three cases of \eq{fourcases}.
At the end of the procedure we will present a summary table collecting
the main results for all the four types of magnetic orders compatible
with factorization.

Given an antiferromagnetic ordering along the
$x$ direction, and keeping in mind that in such case
$\varphi_{\underline{k}}=|\underline{k}|\pi \, \forall
\underline{k}$, \eq{e.magnxy} implies that $M_y=0$ while
$M^x_{\underline{k}} = \pm M_\bot \equiv \pm M_x$ with $M_x=M_\bot
\ge 0$. The $\pm$ sign reflects the fact that the antiferromagnetic
order discriminates between two sublattices, each characterized by
the opposite sign of the staggered magnetization along $x$. Consequently,
the general condition \eq{e.EXE} takes the form
\begin{eqnarray}\label{e.EXEantiferrox}
 0 & =& \left( \sin \theta M_z  - \cos \theta M_x
\right) \times  \nonumber \\
& \times & \left[\cos \theta M_x
 \mathcal{J}^A_x - \sin \theta \left( M_z \mathcal{J}^F_z - h_f
\right) \right] \, ,
\end{eqnarray}
that admits two solutions for $\theta$,
\begin{equation}
\label{e.solutions} \tan \theta = \frac{M_x}{M_z} \; ; \; \; \; \;
\tan \theta = \frac{\mathcal{J}_x^A M_x}{{\mathcal{J}_z^F M_z - h_f}} \; .
\end{equation}
However, since the E-SQUO must be unique \cite{GiampaoloIlluminati}, the two
solutions must coincide. Factorization thus requires that
\begin{equation}\label{e.mz}
M_z = \frac{h_f}{\mathcal{J}_z^F - \mathcal{J}_x^A} \, .
\end{equation}
Moreover, the requirement of a vanishing tangle (single-site entanglement) imposes
\begin{equation}\label{e.mx}
M_x=\frac12\sqrt{1-\frac{4h_f^2}{(\mathcal{J}_z^F - \mathcal{J}_x^A)^2}}\, .
\end{equation}
Combining Eqs.(\ref{e.mz}--\ref{e.mx}) via \eq{e.minimumpoint} yields
a closed expression for the phase $\theta$ as a function of the Hamiltonian
parameters and of the factorizing field $h_f$:
\begin{equation}
\label{e.theta} \cos \theta =  \frac {2
h_f}{\mathcal{J}_z^F - \mathcal{J}_x^A} \; .
\end{equation}
Therefore, \eq{e.theta}, together with the knowledge of
$\varphi_{\underline{k}}=\pi |\underline{k}|$, once inserted into
\eq{e.Ok}, determines, independently of the value of the magnetizations,
the form of the E-SQUO at factorization:
\begin{equation}\label{e.oz}
\bar{O}_k = \cos \theta
S^z_{\underline{i}} + e^{i \varphi_k} \sin\theta S^x_{\underline{i}}\,. \\
\end{equation}
The form of the candidate factorized state $\ket{\Psi_f}$
is then readily determined by imposing that the action of the E-SQUO
on all lattice sites leaves the ground state invariant:
\begin{equation}\label{e.inalterate}
\bigotimes_{\underline{k}} 2 \bar{O}_{\underline{k}} \ket{\Psi_f} =
\ket{\Psi_f} \, .
\end{equation}

>From Eqs.(\ref{e.inalterate}--\ref{e.oz}) we find then that the
candidate factorized ground state $\ket{\Psi_f}$ has to be a tensor
product of single-site states that are the exact eigenstates of
the operators $\bar{O}_{\underline{k}}$ with eigenvalue $1/2$:
\begin{eqnarray}
\label{e.candidate}
\ket{\Psi_f}& = & \bigotimes_{\underline{k}} \ket{\psi_{{\underline{k}}} }\; ;  \\
\ket{\psi_{{\underline{k}}}}&=&
\cos(\theta/2)\ket{\uparrow_{\underline{k}}} +e^{i
\varphi_{\underline{k}} } \sin(\theta/2)
\ket{\downarrow_{\underline{k}}}\; \nonumber .
\end{eqnarray}

Equation (\ref{e.candidate}) expresses the general form that a factorized ground state
must assume given an antiferromagnetic order along $x$. This result is independent of the actual
values assumed by the staggered magnetizations $M_{\alpha}$'s. We remark that in order to
establish the general form of the candidate ground state it is essential to impose the vanishing
of the EXE. In fact, without requiring it, the phase $\theta$ and the candidate factorized
ground state become explicitly dependent on the magnetizations. Then, imposing the conditions
for \eq{e.candidate} to be an eigenstate of the Hamiltonian, one merely obtains a relation
between the factorizing field $h_f$ and the magnetizations $M_\alpha$'s. If the analytic
expression of at least one of the $M_\alpha$'s is known, it is still possible to obtain an
expression of the field $h_f$ in terms of the Hamiltonian parameters. However, this actually
occurs in few special, exactly solvable, cases (for instance, the XY model \cite{Barouch}).
Otherwise, in general one needs to resort to numerical evaluations of the magnetizations or
introduce other approximations in the analysis. This problem is completely eliminated in the
general analytic framework based on the invariance of the factorized ground states under the
action of E-SQUOs and on the vanishing of the EXE at factorization.

Before proceeding further let us clarify a relevant point. Because of the even-odd symmetry
of the energy spectra of the system, as elucidated also in Refs.~\cite{Hoeger,Rossignoli},
the assignment of one of the two different staggered magnetizations to each sublattice is arbitrary.
According to this observation, one sees that, inverting the sign of the magnetization
along the $x$ direction on each site, all the conditions for factorization are still
satisfied but the candidate ground state takes the form
\begin{eqnarray}
\label{e.candidate-1}
\ket{\Psi'_f}& = & \bigotimes_{\underline{k}} \ket{\psi'_{{\underline{k}}} }\; ;  \\
\ket{\psi'_{{\underline{k}}}}&=&
\cos(\theta/2)\ket{\uparrow_{\underline{k}}} +e^{i
\varphi_{\underline{k+1}} } \sin(\theta/2)
\ket{\downarrow_{\underline{k}}}\; \nonumber .
\end{eqnarray}
Obviously the two states in Eq.~(\ref{e.candidate}) and in Eq.~(\ref{e.candidate-1}) are
distinguishable and both $\ket{\Psi_f}$ and $\ket{\Psi'_f}$ are legitimate candidate ground
states of the system, as well as general linear combinations of the two (the latter will be in
general highly entangled). Here and in the following we assume to be working in a situation of
broken symmetry, i.e. after that a small perturbing external field along the $x$ direction
has lifted the degeneracy between $\ket{\Psi_f}$ and $\ket{\Psi'_f}$, e.g. lowering the energy
of the former and rising that of the latter. After a time long enough to ensure convergence to
equilibrium, i.e. relaxation to the state $\ket{\Psi_f}$, the perturbation is switched off: In
the thermodynamic limit, this ensures that the system will remain indefinitely in the state
$\ket{\Psi_f}$.

Having determined the exact form of the candidate factorized ground state, the subsequent step
concerns the determination of the conditions for its occurrence, i.e. the conditions under which
a state of the form \eq{e.candidate} is indeed the eigenstate of $H$, \eq{e.Hamiltonian}, with
the lowest energy eigenvalue. To this aim, taking the value of the external field at the
factorization point, $h=h_f$, it is useful to decompose the total Hamiltonian as a sum of pairwise Hamiltonian terms $H_{{\underline{i}},{\underline{j}}}$, so that
$H|_{h=h_f}=\sum_{{\underline{i}}{\underline{j}}}H_{{\underline{i}}{\underline{j}}}$, where
each term $H_{{\underline{i}},{\underline{j}}}$ involves only the degrees of freedom of a single pair of spins:
\begin{equation}
\label{e.couple}
 H_{{\underline{i}},{\underline{j}}} =
 {J_x^r} S_{\underline{i}}^x S_{\underline{j}}^x  +
 {J_y^r} S_{\underline{i}}^y S_{\underline{j}}^y  +
 {J_z^r} S_{\underline{i}}^z S_{\underline{j}}^z
 - h_f^r \left( {S_{\underline{i}}^z  + S_{\underline{j}}^z }
 \right) .
\end{equation}
Here the quantity $h_f^r$ plays the role of the \virg{component} of the
factorizing field that acts on the selected pair of spins and obeys the
following relation:
\begin{equation}
\label{e.hfij}
2 h_f^r= \cos\theta \left( J_z^r-(-1)^r J_x^r \right)\; .
\end{equation}
The above ensures that $h_f= \sum_r Z_r h_f^r$ satisfies \eq{e.theta}.
Proving that the candidate factorized ground state $\ket{\Psi_f}$ is
an eigenstate of the total Hamiltonian $H$ is equivalent to prove
that it is a simultaneous eigenstate of all pair Hamiltonians
$H_{{\underline{i}}{\underline{j}}}$, or, more precisely, that the
projection of $\ket{\Psi_f}$ onto the subspace of two given spins
$\underline{i}$ and $\underline{j}$ --- which is still a pure
state since $\ket{\Psi_f}$ is a tensor product of single-spin pure
states --- is an eigenstate of $H_{{\underline{i}}{\underline{j}}}$ for
every pair $\{{\underline{i}},{\underline{j}}\}$. To proceed, we
associate to each lattice site a set of orthogonal spin operators defined as
follows:
\begin{eqnarray}
\label{e.axayaz}
 A^x_{\underline{i}} &=& \cos \theta \cos \varphi_{\underline{i}}
        S^x_{\underline{i}}-\sin \theta S^z_{\underline{i}}\,; \nonumber \\
        A^y_{\underline{i}} &=& \cos\varphi_{\underline{i}} S^y_{\underline{i}}\,; \\
        A^z_{\underline{i}} &=& \sin \theta \cos \varphi_{\underline{i}}
        S^x_{\underline{i}}+ \cos \theta S^z_{\underline{i}}\,.
        \nonumber
\end{eqnarray}
It is immediate to observe that $A^z_{\underline{k}}\equiv
\bar{O}^z_{\underline{k}}$. By inverting Eqs.~(\ref{e.axayaz}), we
can conveniently re-express the standard spin operators
as functions of the new set of operators
$\{A^\alpha_{\underline{i}}\}$:
\begin{eqnarray}\label{e.sxsysz}
S_{\underline{i}}^x  &=& \cos\varphi_{\underline{i}} \left(
A_{\underline{i}}^x \cos \theta  + A_{\underline{i}}^z \sin \theta  \right)\,; \nonumber \\
S_{\underline{i}}^y  &=& A_{\underline{i}}^y \cos
\varphi_{\underline{i}}\,; \\
S_{\underline{i}}^z  &=&  - A_{\underline{i}}^x \sin \theta
+A_{\underline{i}}^z \cos  \theta \, . \nonumber
\end{eqnarray}
Inserting \eq{e.sxsysz} in \eq{e.couple} we obtain the expression
of the pair Hamiltonian $H_{\underline{i},\underline{j}}$ as a
function of the sets of operators $\{A^\alpha_{\underline{i}}\}$ and
$\{A^\alpha_{\underline{j}}\}$:
\begin{widetext}
\begin{eqnarray}
\label{e.pairhij} H_{\underline{i},\underline{j}}&=&
A_{\underline{i}}^z A_{\underline{j}}^z \left[  \cos ^2 \theta J_z^r
+ \sin ^2 \theta  J_x^r (-1)^r \right] -2 h_f^r \sin \theta \left[
{A_{\underline{i}}^x \left( A_{\underline{j}}^z -\frac{1}{2}\right)
+\left( A_{\underline{i}}^z-\frac{1}{2}\right) A_{\underline{j}}^x }
\right] \nonumber \\
& & +\ A_{\underline{i}}^x A_{\underline{j}}^x \left( \sin^2 \theta
J_z^r+\cos^2\theta (-1)^r J_x^r \right)+ A_{\underline{i}}^y
A_{\underline{j}}^y J_y^r (-1)^r - h_f^r \cos \theta \left(
{A_{\underline{i}}^z + A_{\underline{j}}^z } \right) \, ,
\end{eqnarray}
\end{widetext}
where we recall that $r$ is the distance between sites $\underline{i}$ and
$\underline{j}$ and that, in the presence of an antiferromagnetic order
along the $x$ direction, $\cos \varphi_{\underline{i}} \cos \varphi_{\underline{j}}=(-1)^r$.
The spin-pair projection $\ket{\psi_{\underline{i}}}\ket{\psi_{\underline{j}}}$ of
$\ket{\Psi_f}$ is an eigenstate of the pair Hamiltonian $H_{\underline{i}\underline{j}}$
either if all terms in \eq{e.pairhij} admit $\ket{\psi_{\underline{i}}}\ket{\psi_{\underline{j}}}$
as an eigenstate or if they annihilate it. By construction, recalling that
$A^z_{\underline{i}}\equiv\bar{O}_{\underline{i}}$ and that
$\ket{\psi_{\underline{i}}}$ is the eigenstate of $A^z_{\underline{i}}$ with eigenvalue equal to $1/2$,
the spin-pair projection of $\ket{\Psi_f}$ is an eigenstate of the pair Hamiltonian if the following condition holds:
\begin{equation}\label{e.condition}
\cos ^2 \theta J_x^r  + (-1)^r \sin ^2 \theta  J_z^r  -J_y^r=0\; .
\end{equation}
For general models of the type \eq{e.Hamiltonian} with a given, arbitrary,
maximum spatial range of interaction $s$, \eq{e.condition} admits the following
solution:
\begin{eqnarray}
\label{e.solcondition}
 \cos ^2 \theta  = \frac{J_y^r - (-1)^r J_z^r}{J_x^r-(-1)^r
J_z^r} \, ,
\end{eqnarray}
for all distances $r \le s$, i.e. for all distances associated with nonvanishing couplings (we
recall that $J_\alpha^r = 0 \ \forall \alpha$ and $\forall r \ge s+1$). The set of equations
(\ref{e.solcondition}) determines the conditions that must be satisfied
simultaneously by the phase $\theta$, which is unique, in order for $\ket{\Psi_f}$
to be an eigenstate of the total Hamiltonian \eq{e.Hamiltonian}. Eqs.~(\ref{e.solcondition})
discriminate quite clearly between short-range models, i.e. models with only one nonvanishing
coupling at $r=s=1$ (models with nearest-neighbor interactions)
and all other possible models containing finite- or infinite-range interaction terms. Specifically,
models with only nearest-neighbor interactions are characterized by the fact that
Eqs.~(\ref{e.solcondition}) reduce to a single condition and hence if the value of the r.h.s. of
Eq.~(\ref{e.solcondition}) falls in the interval $[0,1]$ (associated to the permitted
values of $\cos^2 \theta$ in terms of the interaction parameters), one can immediately
conclude that the models under investigation admit a factorized eigenstate. On the other
hand, for models containing interaction terms of longer range --- so that  Eqs.~(\ref{e.solcondition}) include two or more conditions --- factorized energy eigenstates are allowed if and only if
the r.h.s. of all conditions in the set (\ref{e.solcondition}) take values in the interval $[0,1]$
and, moreover, they all coincide.

Summing term by term over the index $r$ all the relations in \eq{e.condition}, taking into account \eq{e.theta}, and solving for $h_f$, one eventually obtains the exact expression of the
factorizing field as a function of the net interactions:
\begin{equation}
\label{e.factorizingfield}
h_f =  \frac12 \sqrt{(\mathcal{J}^A_x -
\mathcal{J}^F_z ) (\mathcal{J}^A_y - \mathcal{J}^F_z )} \; .
\end{equation}
We remark that \eq{e.factorizingfield}, which was derived in
Ref.~\cite{theory} using some unnecessary auxiliary assumptions,
is completely general and holds for lattices of arbitrary spatial
dimension and for spin-spin interactions of arbitrary range. Accordingly,
the angle $\theta$ that determines the direction of the E-SQUO can be also
expressed as a function of the net interactions as follows
\begin{equation}
\label{e.theta2} \cos \theta =
\sqrt{\frac{\mathcal{J}^A_y-\mathcal{J}^F_z}{\mathcal{J}^A_x-\mathcal{J}^F_z}}\,.
\end{equation}
In the case of a maximum range of interaction $s \geq 2$, relation \eq{e.theta2} for the net
interactions is a necessary but not sufficient condition for factorization, and further
use of Eqs.~(\ref{e.condition}) and Eqs.~(\ref{e.solcondition}) is needed, as we will show
with explicit examples in Sec.~\ref{sec.long} and Sec.~\ref{sec.anisotropy}.

We are finally left with the problem of establishing conditions for a factorized energy eigenstate $\ket{\Psi_f}$ to be indeed a ground state of the system. A very simple sufficient but not necessary condition is that the spin-pair projection of $\ket{\Psi_f}$ onto the subspace of a pair of spins be the simultaneous ground state of every pair Hamiltonian $H_{{\underline{i}}{\underline{j}}}$. In fact, given the relation existing between all the pair Hamiltonians $H_{{\underline{i}}{\underline{j}}}$ in \eq{e.couple} and the total Hamiltonian $H$ \eq{e.Hamiltonian}, it follows that the simultaneous ground state of all pair Hamiltonians $H_{{\underline{i}}{\underline{j}}}$ is also ground state of $H$.
To proceed, let us re-express the pair Hamiltonians \eq{e.pairhij}
in matrix form in the basis spanned by the eigenstates of the operators
$A_{\underline{i}}^z \otimes A_{\underline{j}}^z$. One has
\begin{equation}
\label{e.matrix}
H_{\underline{i}\underline{j}}  = \left( {\begin{array}{*{20}c}
   {\alpha _r  - h_f^r } & 0 & 0 & 0  \\
   0 & { - \alpha _r } & {J_y^r/2 } & { \beta_r }  \\
   0 & {J_y^r/2 } & { - \alpha_r } & { - (-1)^r \beta _r }  \\
   0 & { \beta _r } & { - (-1)^r \beta _r } & {\alpha_r  + h_f^r }  \\
\end{array}} \right) \, ,
\end{equation}
where $\alpha _r  = \frac{1}{4} (\cos ^2 \theta J_z^r  + (-1)^r \sin
^2 \theta  J_x^r)$ and $\beta_r  = h_f^r \sin \theta$. Making use of
\eq{e.theta2} and \eq{e.hfij} the eigenvalues of the matrix \eq{e.matrix}
can be put in the form
\begin{eqnarray}
  \varepsilon_0 &=&  \frac{1}{8} \left[ 3 (-1)^r J_x^r  - J_z^r  -
   \left( {J_z^r-(-1)^r J_x^r  } \right)\cos \left( {2\theta } \right)
  \right] \nonumber \\
  \varepsilon_1 &=& \frac{1}{8} \left[ 3 (-1)^r J_x^r  - J_z^r  -
  \left( { J_z^r -(-1)^r J_x^r} \right)\cos \left( {2\theta } \right)
  \right] \nonumber \\
  \varepsilon_2 &=& \frac{1}{8} \left[ - 3\left( {(-1)^rJ_x^r  + J_z^r } \right)
   + \left( {J_z^r-(-1)^r J_x^r  } \right)\cos \left( {2\theta } \right)
  \right]\nonumber \\
  \varepsilon_3 &=& \frac{1}{8} \left[ { - 3(-1)^rJ_x^r  + 5J_z^r  +
  \left( {J_z^r- (-1)^r J_x^r} \right)\cos \left( {2\theta } \right)}
 \right] \nonumber \\
 &&
\end{eqnarray}

The eigenvalue $\varepsilon_0$ is the one associated to the
projection $\ket{\psi_{\underline{i}}}\ket{\psi_{\underline{j}}}$ of
the factorized eigenstate $\ket{\Psi_f}$ onto the four-dimensional
Hilbert space associated to the pair of spins $\{ S_{\underline{i}}, S_{\underline{j}} \}$.
Therefore, the spin-pair projection is the ground state of the pair Hamiltonian if
$\varepsilon_0 \le
\{\varepsilon_1,\,\varepsilon_2,\,\varepsilon_3\}$. Imposing this condition yields
the following inequalities:
\begin{eqnarray}
 (J_x^r+J_y^r) \left[ J_x^r - (-1)^r J_z^r\right] &\ge& 0 \; ; \label{e.suffcondition.1} \\
(-1)^r J_x^r - J_z^r & \le & 0 \; . \label{e.suffcondition.2}
\end{eqnarray}
Notably, these inequalities are automatically satisfied for any interacting spin system that is not frustrated and that verifies all of the Eqs.~(\ref{e.solcondition}). Namely, for every even value of the distance $r$ in the interaction range ($r \le s$) \eq{e.suffcondition.2} implies that the denominator in the corresponding \eq{e.solcondition} is non-positive. Therefore, an acceptable value of $\cos^2 \theta$ can be obtained only if also the numerator is non-positive and exceeds or equals the denominator. These conditions in turn imply the order relation, $J_x^r \le J_y^r$. On the other hand, comparing \eq{e.suffcondition.2} and \eq{e.suffcondition.1}, one has that $J_x^r \le -J_y^r$. Therefore, comparing the two order relations yields finally $J_x^r \le - |J_y^r|$  ($r$ even). This condition is always verified in systems with an antiferromagnetic order along the $x$ direction and in absence of frustration and therefore the projection of the factorized energy eigenstate on the Hilbert space of every pair of spins is indeed the ground state of every pair Hamiltonian. On the other hand, in the case of
odd $r$, \eq{e.suffcondition.2} implies that the denominator of
\eq{e.solcondition} is non-negative and, similarly to the
former case, the numerator must be non-negative and not exceeding
the denominator. Hence we obtain $J_x^r \ge J_y^r$ while from
\eq{e.suffcondition.1} we recover $J_x^r \ge -J_y^r$ that, in turn, implies
$J_x^r \ge |J_y^r| $. Again, such inequalities are always verified
in the presence of an antiferromagnetic order and in the absence of
frustration.
\begin{center}
\begin{table*}[]
\centering
\caption{Summary of results for the four magnetic orderings [\eq{fourcases}]
compatible with ground state factorization. \\ \quad \\}
\label{t.summary}
\begin{tabular}{cccc}
\hline \hline
&&& \\
magnetic order & $\cos^2\theta$ & factorizing field $h_f$
& single-spin state $\ket{\psi_{{\underline{k}}}}$ \\
&&& \\ \hline

&&& \\
$\mu=\mathcal{J}_x^F$ \; & $  \frac{J_y^r
-J_z^r}{J_x^r-J_z^r}$ \; \; & \; $\frac{1}{2} \sqrt{(\mathcal{J}^F_x -
\mathcal{J}^F_z ) (\mathcal{J}^F_y - \mathcal{J}^F_z )}$ \; \; &
\; $\cos(\theta/2)\ket{\uparrow_{\underline{k}}} + \sin(\theta/2)
\ket{\downarrow_{\underline{k}}}$ \\
&&& \\ \hline

&&& \\
$\mu=\mathcal{J}_x^A$ & $  \frac{J_y^r -(-1)^r
J_z^r}{J_x^r - (-1)^r J_z^r}$ & $ \frac{1}{2}
\sqrt{(\mathcal{J}^A_x - \mathcal{J}^F_z ) (\mathcal{J}^A_y -
\mathcal{J}^F_z )}$ & $ \cos(\theta/2)
\ket{\uparrow_{\underline{k}}} + (-1)^{|\underline{k}|}
\sin(\theta/2)
\ket{\downarrow_{\underline{k}}}$ \\
&&& \\ \hline

&&& \\
$\mu=\mathcal{J}_y^F$ & $  \frac{J_x^r
-J_z^r}{J_y^r-J_z^r}$ & $\frac{1}{2} \sqrt{(\mathcal{J}^F_x -
\mathcal{J}^F_z ) (\mathcal{J}^F_y - \mathcal{J}^F_z )}$ &
$\cos(\theta/2)\ket{\uparrow_{\underline{k}}} + i \sin(\theta/2)
\ket{\downarrow_{\underline{k}}}$ \\
&&& \\ \hline

&&& \\
$\mu=\mathcal{J}_y^A$ & $  \frac{J_x^r -(-1)^r
J_z^r} {J_y^r- (-1)^r J_z^r}$ & $ \frac{1}{2}
\sqrt{(\mathcal{J}^A_x - \mathcal{J}^F_z ) (\mathcal{J}^A_y -
\mathcal{J}^F_z )}$ & $ \cos(\theta/2)
\ket{\uparrow_{\underline{k}}} + (-1)^{|\underline{k}|} i
\sin(\theta/2)
\ket{\downarrow_{\underline{k}}}$ \\
&&& \\ \hline \hline
 \end{tabular}
\end{table*}
\end{center}

Collecting all these results, we have proved that the following holds:

\medskip

\noindent {\bf Theorem}.
{\it For any cooperative system of spin-$1/2$ particles described by Hamiltonians $H$
of the type [\eq{e.Hamiltonian}], characterized by an anti-ferromagnetic order along the $x$ axis as emerging from \eq{fourcases} and in the absence of frustration, the simultaneous
verification of all Eqs.~(\ref{e.condition}) is necessary and sufficient for the fully
factorized state \eq{e.candidate} to be the exact ground state of $H$ when the external
magnetic field takes the value $h=h_f$ determined by \eq{e.factorizingfield}}.

\medskip

This central result holds as well, with obvious modifications,
when one considers the other different ordered phases that form
the set identified in \eq{fourcases}. In Tab.~\ref{t.summary} we
summarize and compare results for each of the four different
possibilities.

\subsection{Factorization, balancing, and saturation}\label{s.sec.saturation}

Before moving to apply the general method to specific models and examples
of conceptual and physical relevance, we discuss briefly the meaning of
factorization and its marked differences with the phenomenon of ``saturation"
(see below). As we have seen, and as it will appear even clearer in the
discussion of the examples in the following sections, the physical mechanism
of ground state factorization in quantum spin systems is due to a delicate kind
of ``balancing" between the coupling strengths that regulate the intensity
(and range) of the interactions and the aligning effect of the external field,
that tends to orient all spins along a given direction and to destroy all
quantum correlations. The remarkable aspect of factorization is then
that it occurs, according to well defined conditions and constraints, at precisely
defined, finite values of the couplings and of the fields in an ordered
(or symmetry-broken) phase. This means that factorization occurs when the
system is relatively strongly interacting and the external aligning field is
relatively weak. Moreover, we will see that the factorization point is always a
precursor from below (from the ordered phase) of the quantum critical point,
as first observed for the $XYZ$ model with nearest-neighbor interactions \cite{Roscilde}.

Saturation is the phenomenon of ground state factorization that occurs
trivially when the value of the external field grows unboundedly compared
to all other Hamiltonian parameters. Therefore its aligning effect on the spins
prevails against all other effects and quantum correlations are suppressed.
The main physical difference between proper factorization and saturation is then
clear and is reflected at the level of the Hilbert space of states. Namely,
suppose that the system under study admits a fully factorized state that
is the true ground state of the system at $h=h_f$. Then, when one moves
away from the factorizing field, i.e. when $h-h_f=\epsilon\neq0$, the
factorized state ceases to be an eigenstate of the Hamiltonian. On the
contrary, in the presence of saturation, the trivially
ensuing factorized state is always an eigenstate of the Hamiltonian and remains
the ground state of the system when perturbing the value of the external field $h$.

Let us formalize the above discussion. Consider a system that
admits a fully factorized ground state. In this case the pair
Hamiltonian in \eq{e.pairhij} admits as eigenstate a factorized
spin-pair state. However, \eq{e.pairhij} is the expression of the
pair Hamiltonian only at $h=h_f$. If we consider a value of the
external field different from the factorizing field we must include
in \eq{e.pairhij} a term proportional to
$(A_{\underline{i}}^x-A_{\underline{j}}^x) \sin \theta (h-h_f)$.
Obviously, for nonvanishing values of $\sin \theta$ the presence of such
term prevents the factorized state from being an eigenstate of the
Hamiltonian of the system. On the contrary, if $\sin \theta=0$ and hence if
$\bar{O}_{\underline{k}} \equiv S_{\underline{k}}^z \, \forall
\underline{k}$ the extra Hamiltonian term vanishes and the factorized state
$\bigotimes_{\underline{k}}\ket{\uparrow}_{\underline{k}}$ remains
an energy eigenstate of the system for all values of the external field.
This result can be understood from a different point of view. When
$\theta=0$, from Eqs.(\ref{e.solcondition}) we immediately obtain
that $J_x^r=J_y^r$, which implies that the total magnetization along
the $z$ axis is preserved,
$[H,\sum_{\underline{k}}S^z_{\underline{k}}]=0$. Therefore the
Hamiltonian of the system and the total magnetization operator admit
a complete set of common eigenstates, discriminated by the
eigenvalue of the total magnetization: the factorized state is the
one (and only one) characterized by the maximum eigenvalue.

We now move on to show that the factorized eigenstate, in the case of
saturation, is the system's ground state in an infinite range of values of
the external field. Here and in the following we recall that
since the external field is directed along the conventional $z$ direction,
only saturation with spin alignment along the $z$-axis is allowed.
At very low values of $h$ the ground state of the system will
possess a small total magnetization along the field direction.
However, as $h$ increases, also the total ground-state magnetization
along the $z$ direction will increase until all spins align with the field
and saturation occurs. By increasing $h$ further, it is not
possible for the system to evolve in a state with higher values of the magnetization
along the field direction. Hence, the factorized state remains the ground
state of the system for all values of the external field $h \ge h_s$, where
by $h_s$ we denote the value of the field for which the system reaches
saturation (``saturating field''). Let us consider the matrix representation, similarly to
Eq. (\ref{e.matrix}), of the pair Hamiltonian at arbitrary values of $h$. Taking
into account the condition $\sin \theta=0$ we have
\begin{equation}\label{e.matrix2}
H_{\underline{i}\underline{j}}  = \left( {\begin{array}{*{20}c}
   {J_z^r/4  - h^r } & 0 & 0 & 0  \\
   0 & { - J_z^r/4 _r } & {J_y^r/2 } & { 0 }  \\
   0 & {J_y^r/2 } & {- J_z^r/4 _r } & {0 }  \\
   0 & { 0 } & { 0} & { J_z^r/4 _r  + h_f^r }  \\
\end{array}} \right) \, .
\end{equation}
It is immediate to verify that if $h^r \ge |J_x^r|/2+J_z^r/2 = h^r_s$, the
projection of the factorized state onto the considered pair of
spins is associated to the lowest energy. Re-summing over all
values of the spatial range index $r$ and taking into account the coordination
number $Z_r$ of the lattice, we have that the factorized state is
the ground state of the total system if
\begin{equation}
\label{e.hfsaturation}
h \ge h_s =\frac{1}{2} \left( \sum_r Z_r |J_x^r| + \sum_r Z_rJ_z^r
\right) \, .
\end{equation}
Eq.~(\ref{e.hfsaturation}) is the rigorous condition for saturation. We notice that
it coincides with \eq{e.factorizingfield} in the
limit $J_y^r\rightarrow J_x^r$, meaning that in this simple instance
no ground-state factorization occurs other than saturation.

\bigskip

In the following Sections we apply the general theory
of factorization to different quantum spin models. We will
derive a series of exact results concerning the occurrence
of factorized ground states in systems with various types of
two-body interactions of different spatial ranges defined on
regular lattices of arbitrary spatial dimensionality.
The general method enables to determine many novel exact
factorized ground state solutions as well as the existence
and the nature of ordered phases in models that are, in general,
non exactly solvable. Moreover, it allows to recover quite
straightforwardly the existing analytical and numerical
results. In the process, we will come to appreciate that
ground state factorization is a phenomenon more common than
previously believed. Determining the existence of exact factorized
solutions in non exactly solvable models should in turn allow to
envisage controlled approximation schemes, e.g. perturbative or variational,
to investigate the physics of non exactly solvable quantum spin models
in the vicinity of quantum factorization points.

\section{Models with short-range interactions} \label{sec.short}

\begin{figure}[tb]
\centering
\includegraphics[width=8.5cm]{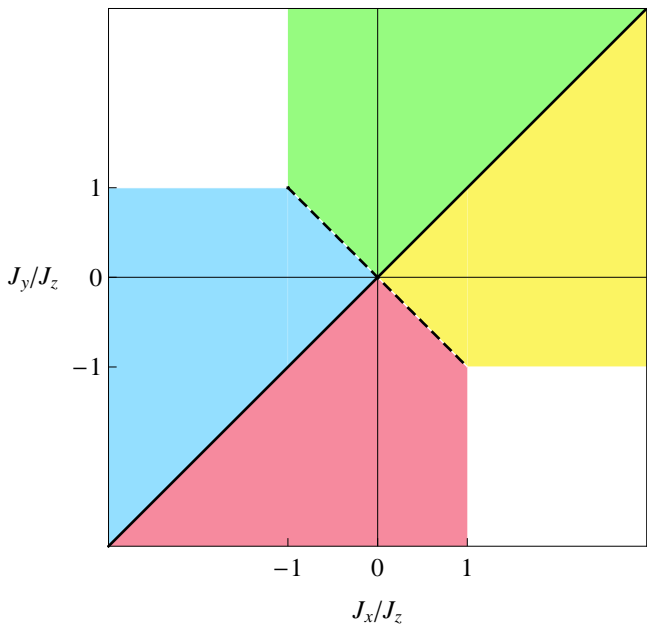}
\caption{(Color online) Diagram of ground state factorization as a function of the ratios $J_x/J_z$
and $J_y/J_z$. The coupling $J_z$ is set to be positive for reference. Analogous
results hold in the case $J_z < 0$. Blank regions: No factorization allowed.
Rightmost (yellow online) shaded region: Presence of a fully factorized ground
state supporting an antiferromagnetic order along the $x$ direction.
Leftmost (light blue online) shaded region: Fully factorized ground state
with a ferromagnetic order along $x$. Bottommost (light red online) shaded region:
Fully factorized ground state with a ferromagnetic order along the $y$ direction.
Topmost (light green online) shaded region: Fully factorized ground state with an
antiferromagnetic order along $y$. The solid line $J_x=J_y$ corresponds to models
that exhibit saturation rather than true factorization; the dashed line $J_y=-J_x$
accommodates for models with unresolved superpositions of two fully factorized ground
states each one supporting a different magnetic order. Either one of the two possible ground
states is then "chosen" by the system depending on the way the symmetry is broken, see
text for further details. All plotted quantities are dimensionless.}
\label{f.factor}
\end{figure}

In this and in the following sections we will consider quantum spin
models of the $XYZ$ type \eq{e.Hamiltonian} defined on regular lattices
of arbitrary dimension. Let us first consider models with short-range
interactions including only nearest-neighbor couplings:
\begin{equation} \label{exe.cond.1}
J_\alpha^1\equiv
J_\alpha,\,\quad J_\alpha^r=0\,, \forall r \ge 2.
\end{equation}
As in Sec. \ref{sec.Method} we restrict ourselves to the case in
which such models are defined on regular lattices of a generic
dimension, whose geometry does not induce any frustration, i.e. in
which any closed loop is characterized by an even number of spins.
The absence of geometrical frustration together with the fact that
all the couplings beyond the nearest neighbor spins vanish, ensure
that it is always possible for the system to satisfy all the conditions
on the net magnetic interactions. Therefore, as shown in
Sec.~\ref{sec.Method}, if the r.h.s. of \eq{e.solcondition} --- or, better,
if the r.h.s. of the equation relative to each possible magnetic
order (See Tab. \ref{t.summary}) takes an acceptable value
($0\le\mathrm{ r.h.s.} \le 1$) --- then the system admits
a fully factorized ground state. In the case of
antiferromagnetic ordering along the $x$ direction, which
is obtained if $J_x \ge |J_y|$, from \eq{e.solcondition} it follows
that a factorized ground state exists if $J_y \le -J_z$. Analogous conditions
are obtained for the other types of magnetic orders compatible with full
ground state separability. The results are summarized in Fig. \ref{f.factor},
where $J_z$ is set assume positive values for reference. A very similar diagram
of factorization holds in the case $J_z < 0$.

As one can see from Fig. \ref{f.factor}, the type of magnetic order
corresponding to the factorized ground state determines four different
regions in the space of the Hamiltonian parameters (one for each
different ordering) associated to the existence of a factorized ground
state. The four different domains of the interaction parameters
are separated by two bisectrices that individuate special cases.
The line $J_x=J_y$ is constituted by the set of points in which
saturation occurs in place of genuine factorization (see SubSec.~\ref{s.sec.saturation}).
On the other hand, the line $J_x=-J_y, \; -1 \le J_x/J_z \le 1$,
corresponds to models that, for a given value of the factorizing field $h_f$
and of the E-SQUO orientation $\theta$, admit two degenerate, but distinct
factorized states. The choice of the ground state is then determined by the
usual mechanisms of symmetry breaking: If an infinitesimal magnetic field
is applied in the $xy$ plane, the axis of magnetic alignment in the ground state
is determined by the order in which the two limits $h_x \rightarrow 0$ and $h_y \rightarrow 0$
are realized.

It is important to observe that all the shaded regions of Fig.~\ref{f.factor}
correspond to exact analytical ground-state solutions
of the investigated short-range models, which are, in general, non-exactly
solvable. The exact, fully factorized ground states are tensor products
of the form reported in the fourth column of Tab. \ref{t.summary}, in which
the value of the orientation parameter $\theta$ of the E-SQUO is retrieved
by solving the equation in the second column of Tab. \ref{t.summary}.
Obviously, such an exact solution holds only at the factorizing field $h=h_f$.
In the case of antiferromagnetic order either along the $x$ or the $y$ direction,
the latter is given by (See Tab. \ref{t.summary})
\begin{equation}\label{e.sh.hfanti}
h_f = \frac{Z_1}{2}\sqrt {\left( {J_x  + J_z } \right)\left( {J_y +
J_z } \right)} \; ,
\end{equation}
while in the case of a ferromagnetic phase, again unrespectfully if
along $x$ or along $y$, it reads
\begin{equation}\label{e.sh.hfferro}
h_f = \frac{Z_1}{2}\sqrt {\left( {J_x  - J_z } \right)\left( {J_y
-J_z } \right)} \; .
\end{equation}
Given that the ground state is known exactly at the factorization point,
all the relevant physical observables can be evaluated straightforwardly.
For instance, the ground state energy density at factorization reads
\begin{equation}\label{e.sh.Eanti}
\frac{E}{N} = - \frac{Z_1}{8}\left( {J_x  + J_y  + J_z } \right)
\end{equation}
in the presence of an antiferromagnetic ordering,
while in the ferromagnetic case it is
\begin{equation}\label{e.sh.Eferro}
 \frac{E}{N} =  \frac{Z_1}{8}\left( {J_x  + J_y  -J_z } \right)\, .
\end{equation}

Both the factorizing field and the ground state energy, unlike
the E-SQUO orientation $\theta$ and the expression of the factorized
ground states, are functions of the number of nearest neighbors per
site (the coordination number $Z_1$, where the index $1$ denotes the
nearest-neighbor range of the interactions). Hence, they depend on the
geometry and the spatial dimension of the lattice. To begin with,
let us consider one-dimensional lattices (spin chains, $D=1$), for which
the coordination number $Z_1=2$. Substituting this value in the previous
equations, we immediately recover, among others, the value of the factorizing field
and the ground state energy obtained analytically by Kurmann {\em et
al.} for the $XYZ$ spin chain using the direct method \cite{Kurmann}.
Kurmann {\em et al.} had to restrict their investigation to coupling
constants $J_\alpha$ taking values in the interval $[0,1]$, thus
establishing ground state factorization only in a restricted set of
conditions. Instead, the general method allows to assess the occurrence
of factorized ground states associated to different types of magnetic
orderings in the entire region of the Hamiltonian parameters.

One of the most important advantages of the general method is that it
is easily applicable irrespective of the spatial dimensionality, while
the direct method becomes increasingly difficult to apply with growing
spatial dimensions of the lattice. Numerical techniques can overcome some
of the limits of the direct method, although they fall short of the power and
extension of the general analytic method. The most important numerical result
has been obtained by applying quantum Monte Carlo techniques to the study
of the $XYZ$ model in $D=2$ space dimensions (square lattice) \cite{Roscilde}.
Applying the present method to the $XYZ$ model on the square lattice allows
not only to immediately reobtain analytically the previous numerical findings,
but also, again, to extend the investigation to the entire space of the
Hamiltonian parameters by simply taking into account that for the square lattice
$Z_1=4$. Moreover, the flexibility of the general method allows to include in the analysis
different planar models. For instance, considering models defined on planar hexagonal lattices,
the previous analysis needs to be modified only in that the value of the coordination number,
in this case is $Z_1=3$.

Let us now move to $D = 3$ and investigate ground state factorization in nearest-neighbor
$XYZ$ models on three-dimensional regular lattices, for which there were so far no results
(apart the preliminary ones obtained in our previous work \cite{theory}) either analytic
or numerical (such a task would be extremely demanding if approached either via the direct
method or via numerical techniques). Resorting to the general method, we can instead fully
characterize the factorized ground state and the associated physical quantities just by
entering the correct value of the coordination number, e.g. $Z_1=6$ for a simple cubic lattice.
Obviously, the results extend straightforwardly also for models defined on lattices of arbitrary
higher dimension $D > 3$.

\section{Models with finite- and long-range interactions} \label{sec.long}

The analytic method can be applied with minor complications to models with
interactions of longer spatial range. As already mentioned in
Sec.~\ref{sec.Method}, concerning ground state factorization,
the main difference between models with short- and finite-range
interactions is that for the latter there exist more than one
equation in the set (\ref{e.solcondition}). Each of these equations
is associated to a different value of the range $r \le s$, where $s$
is the maximum interaction range, that is the maximum distance between
two directly interacting spins: All the couplings vanish for pairs
of spins with inter-spin distance greater or equal to $s+1$.

Clearly, the fact that all the equations have to be solved
simultaneously yields tighter conditions on factorizability
compared to the case of models with nearest-neighbor interactions.
Namely, it is not sufficient that the r.h.s. of each Eq.~(\ref{e.solcondition})
takes values in the interval $[0,1]$. It is also necessary to require
that all these values coincide. This constraint yields a set
of further conditions that must be satisfied by all the nonvanishing
couplings. For instance, let us suppose that the system is in an antiferromagnetic
ordered phase along the $x$ direction. Labeling by $c$ the value of $\cos^2 \theta$
and imposing that for every $r$ the r.h.s. of Eqs.(\ref{e.solcondition})
must equate $c$, we find that
\begin{equation}\label{e.exe.cantix}
J_y^r=c J_x^r+(-1)^r (1-c) J_z^r \; \; \; \forall r \le s \; ,
\end{equation}
where, to ensure the absence of frustration, $J_x^r$ must obey the
following inequality: $(-1)^r J_x^r \le  J_z^r (c-1)/(1+c) \; \forall
r \le s$. Analogously, for an antiferromagnetic order along $y$:
\begin{equation}\label{e.exe.cantiy}
J_x^r=c J_y^r+(-1)^r (1-c) J_z^r \; \; \;  \forall r \le s \; ,
\end{equation}
and to exclude frustration we must also have that $(-1)^r J_y^r \le J_z^r
(c-1)(1+c) \; \forall r \le s $. In the presence of a
ferromagnetic phase, respectively for an ordering along
$x$ and along $y$, the constraints and the no-frustration condition become
\begin{eqnarray}
J_y^r=c J_x^r+ (1-c) J_z^r & &  J_x^r \le
\frac{c-1}{1+c} J_z^r \; \; \; \forall r \le s \; ; \label{e.exe.cferrox} \\
J_x^r=c J_x^r+(1-c) J_z^r & &  J_y^r \le
\frac{c-1}{1+c} J_z^r   \; \; \; \forall r \le s \; . \label{e.exe.cferroy}
\end{eqnarray}
Eqs. (\ref{e.exe.cantix}--\ref{e.exe.cferroy}) generalize the
instance analyzed in Ref. \cite{theory}, in which we limited the
analysis to the case of vanishing $J_z$ that, according to Eqs.
(\ref{e.exe.cantix}--\ref{e.exe.cferroy}), implies $c=J_x^r/J_y^r$
for all $r \le s$.

Especially in the presence of a ferromagnetic order in the lattice,
Eqs.(\ref{e.exe.cferrox}--\ref{e.exe.cferroy}) allow us to simplify
the expressions of both the factorizing field and the
energy per site. Namely, taking into account the definition of the
ferromagnetic net interactions \eq{e.netinteractions} and the
expression of the factorizing field in the presence of a ferromagnetic
order (see Tab. \ref{t.summary}), we have
\begin{eqnarray}\label{e.exe.hffinit}
h_f=\sqrt{c}\left(\mathcal{J}_x^F-\mathcal{J}_z^F\right) \; , \nonumber \\
h_f=\sqrt{c}\left(\mathcal{J}_y^F-\mathcal{J}_z^F \right) \; ,
\end{eqnarray}
respectively for an order along the $x$ and the $y$ directions. Similarly,
for the energy density we obtain
\begin{eqnarray}\label{e.exe.Efinit}
\frac{E}{N}=(1+c) \mathcal{J}_x^F-c \mathcal{J}_z^F\, , \nonumber \\
\frac{E}{N}=(1+c) \mathcal{J}_y^F-c \mathcal{J}_z^F\, .
\end{eqnarray}

We now discuss a specific example (out of many possible ones) which demonstrates the power of the analytic method in conditions where the more conventional direct approach based on the factorized ansatz of Refs.~\cite{Kurmann,Rossignoli} is completely ineffective without the supplement of a remarkable intuition in guessing the correct forms of the possible factorized ground states.
Let us consider a $XYZ$-type spin model defined on a spatially isotropic cubic lattice of arbitrary size, possessing nearest-neighbor and next-nearest-neighbor interactions, and admitting arbitrarily different couplings along the $x$, $y$, and $z$ directions. The coordination numbers associated to the two types of interaction can be straightforwardly computed from the geometry of the problem and are $Z_1=6$ and $Z_2=18$.  Suppose the model is deduced from an explicit experimental realization of a spin lattice and we are provided with explicit values of the various interaction strengths. For example,
$J_x^1=1, \, J_y^1=0.3, \, J_z^1=0.4, \, J_x^2=-0.6, \, J_y^2=-0.25, \, J_z^2=0.1$.
The application of our ``on-demand'' ground-state factorization search engine is now immediate. From \eq{fourcases} we have that the candidate factorized state must possess antiferromagnetic order along
the $x$ direction. Therefore the general conditions for the existence of a factorization point, as determined by \eq{e.solcondition}, can be summarized as follows
\begin{equation}\label{e.cubicex.condom}
\cos^2\theta=\frac{J_y^1+J_z^1}{J_x^1+J_z^1} = \frac{J_y^2-J_z^2}{J_y^2-J_z^2}
= \frac{J_y^1+J_z^1-3(J_y^2-J_z^2)}{J_x^1+J_z^1-3(J_x^2-J_z^2)} \, .
\end{equation}
This condition must hold together with the frustration-free constraint.
It is readily verified that $\cos^2\theta=0.5$ provides a solution to the  combined set of necessary and sufficient conditions for ground state factorization. Therefore if the experimentalist tunes the external field $h$ at the factorizing value $h_f = \simeq 7.425$, obtained from \eq{e.factorizingfield}, the engineered cubic spin lattice system will relax in a completely factorized ground state of the form given in Table \ref{t.summary} (second row, last column) with $\theta=\arccos(\sqrt{0.5})$.

Many other applications can be considered to models with arbitrary spin-spin interactions of longer range, and they can be treated with the same simplicity within the framework of the general analytic method. For instance, a very interesting example of a model with infinite-range interactions will be treated in the next Section. However, before moving to the study of this case, we should comment on the important issue of the interplay between factorization and frustration in quantum spin systems with multiple spatial scales of interaction.

The investigation of general models including many interaction terms of different spatial ranges
needs to be carried out with particular care, because the interplay between the different interactions
can lead in general to important effects of frustration that tend to suppress
the occurrence of ground state factorization. The very delicate and tricky nature
of the problem is at the origin of a recent incorrect prediction about the existence
of factorized ground states in one-dimensional $XYZ$ models on finite rings
with long-range interactions, obtained exploiting the direct method \cite{Giorgi}.
In Ref. \cite{Giorgi} the author, resorting to the direct method, concludes
that all the factorized energy eigenstates in one-dimensional long-range $XYZ$ models,
with interaction terms of spatial ranges that are in integer ratio
with the total number of sites, are also the states of lowest energy.
However, the proof of this statement obtained using only the direct method requires
that there must be no frustration in the system \cite{Kurmann}. Unfortunately,
the interplay between the interactions of different spatial range can in general
introduce frustration effects (e.g. in the case of antiferromagnetic nearest-neighbor
and next-to-nearest neighbor interactions, and so on). The existence of these effects 
nullifies all proofs, based on the direct method, of the existence of fully separable 
ground states. In fact, immediate counterexamples can be given of fully factorized 
energy eigenstates that are not ground states. Let us consider for instance an $XY$ 
model on a finite ring of six spins with nearest-neighbor and next-to-nearest-neighbor 
antiferromagnetic interactions, and $J_x^1 = 1, \, J_y^1 = 0.3, \, J_x^2 = 0.6, \, J_y^2 = 0.18$.
Following Ref. \cite{Giorgi}, this model should admit a factorized
ground state. In fact, a factorized energy eigenstate with antiferromagnetic order does
indeed exist, with an energy $E=-0.78$. However, solving the model by exact diagonalization
yields that the lowest energy is $E \simeq -1.11$ and that the associated ground state is entangled.
Essentially the same type of counterarguments nullifies a recent claim that the factorized
energy eigenstates are the factorized ground states of mixed-spin models of ferrimagnetism
of the $XYZ$ type \cite{Langari}. A general analysis of ground state factorization in 
ferrimagnetic models with mixed spin-$1/2$-spin-$1$ interactions will be presented elsewhere
\cite{Inprep}, based on a proper extension of the formalism of SQUOs to
include spin-$1$ systems, as sketched in the final part of Ref. \cite{GiampaoloIlluminati}.

In conclusion, the study of factorization in quantum spin models with many different
finite- and long-range interaction terms requires a very careful analysis
and the use of the general analytic method whenever frustration effects are present.
A rigorous and systematic study of the crucial (and subtle) interplay between frustration
and factorization in quantum spin systems will be the subject of a forthcoming paper \cite{Inprep2}.

\section{Lipkin-Meshkov-Glick model}\label{sec.lmg}

A very interesting limit of the cases analyzed in
SubSec.~\ref{sec.long} is given by models with infinite range
interactions, such as the  fully connected or Lipkin-Meshkov-Glick
(LMG) model \cite{Lipkin}, extensively studied in condensed matter physics
\cite{Dusuel,Ivette,Fazio}, which is described by the following Hamiltonian
\begin{equation}\label{e.exe.HLMG}
H= - \frac{1}{4(N-1)} \sum_{i,j} \left( S_i^x S_j^x+ \Delta\  S_i^y
S_j^y \right) -  \frac{h}{2} \sum_i S_i^z \; ,
\end{equation}
with $0\le \Delta \le 1$. The pre-factor $1/4(N-1)$ (with $N$ denoting
the total number of spins) ensures the linear divergence of the ground state energy
as $N$ increases. Let us remark that normalizing the interaction part of the
Lipkin-Meshkov-Glick Hamiltonian requires some care: Being a connected interaction,
it scales as $N \times (N-1)$, i.e. as the product of the total number of
spins and the number of pairs of spins. Therefore, to ensure that
the interaction energy is extensive, i.e. scales with $N$, the
interaction part of the Hamiltonian must be normalized by $N-1$.
Usually, for calculational ease, the interaction part is normalized
by $N$, so that it scales as $N-1$. Using this convention is strictly
speaking incorrect, but is clearly harmless for large $N$ and in the
thermodynamic limit. For instance, concerning the separability of the
ground state, it introduces a slight, spurious $N$-dependence of the
factorization point through the multiplicative factor $(N-1)/N$, and
the error vanishes in the thermodynamic limit. However, on fundamental
physical grounds, the factorization points of translationally invariant
systems, if they exist, must be independent of the size of the system,
and  must take the same value both at finite $N$ and in the thermodynamic
limit. Applying the Kurmann-Thomas-M\"uller factorized ansatz \cite{Kurmann} and the $1/N$ normalization,
Dusuel and Vidal obtained an analytic expression of the factorizing field for the LMG model
that is $N$-dependent through the multiplicative factor $(N-1)/N$ \cite{Dusuel}. Although
defined according to an anomalous convention, this result is essentially exact in the limit
of large $N$.

We first observe that the LMG model in \eq{e.exe.HLMG} can be obtained as
the limiting case of a generic model described by \eq{e.Hamiltonian} assuming
that the range of interactions diverges (i.e $s \rightarrow \infty$) while
the individual strengths become independent of the distance ($J_\alpha^r=J_\alpha$),
scale with $N$ ($J_\alpha \propto 1/N$), and vanish along the $z$-direction
($J_z=0$). Therefore, assuming $J_{x}^r=-2/N$, $J_{y}^r=-2\Delta/N$
and $J_z=0$ in \eq{e.Hamiltonian} we recover \eq{e.exe.HLMG}.

Because of the sign of the interactions and of the range of values
that can be assumed by $\Delta$, we immediately obtain that the
energy per site (\ref{e.energy}) is minimized by a state possessing
ferromagnetic order along the $x$ direction. Moreover, taking into account
that the coupling strengths are proportional to the ratio $1/N$, we have
that the net interactions converge to finite values in the
thermodynamic limit: $\mathcal{J}_{x}^F \rightarrow - 2$,
$\mathcal{J}_{y}^F \rightarrow - 2 \Delta$. It is then rather
straightforward to apply the analytic method to the LMG model and hence
prove, exactly, that the ground state is fully factorized at the
factorizing field $h_f=\sqrt{\Delta}$, which is, correctly, independent of $N$.
The form of the fully separable ground state is the tensor product of
local states of the form $\ket{\psi_k}=\cos
(\theta/2) \ket{\uparrow_k}+\sin (\theta/2) \ket{\downarrow_k}$ with
$\theta=\arccos(\sqrt{\Delta})$ and the energy density reads
$E/N = -(1 + \Delta)/4$. These results are in agreement with those obtained, using
the direct method, by Dusuel and Vidal \cite{Dusuel}, modulo a proper normalization factor.

\section{Models with spatial anisotropies} \label{sec.anisotropy}

The versatility and generality of the analytic method can be demonstrated
further by considering extensions of the translationally-invariant models
discussed in the previous Sections. In fact, till now, we have always
considered systems in which the coupling amplitudes depend only on
the distance between the spins involved. In the present subsection
we shall focus instead on models which contemplate the possibility
of spatial anisotropies, i.e. given the spin $S_{\underline{i}}$
associated to site $\underline{i}$, the interaction coupling with spin
$S_{\underline{j}}$ associated to site $\underline{j}$ depends not only
on the distance $r = |\underline{i}-\underline{j}|$ but also on the
location of site $\underline{j}$ relative to site $\underline{i}$.

To simplify the notations, the physical analysis, and the number of possible
cases that need to be considered, we limit ourselves to
models with short-range interactions, i.e. with all the couplings vanishing
for $r = |\underline{i}-\underline{j}| \ge 2$. We will comment briefly on
general models with spatial anisotropies and interactions of arbitrary range
at the end of this Section. Let us assume that for each site $\underline{i}$
the total number of nearest neighbors is $Z_1 = Z_1^{(1)} + Z_1^{(2)}$, where
$Z_1^{(1)}$ is the number of nearest neighbors whose coupling strength with
$S_{\underline{i}}$ takes a certain value $f_\alpha$, and $Z_1^{(2)}$ is the number of
nearest neighbors whose coupling with $S_{\underline{i}}$ is given by a different value
$g_\alpha$. Obviously, we could consider situations more complicated at will, with
an arbitrary number of $n$ different types of nearest neighbors. For the given example,
the system Hamiltonian reads
\begin{eqnarray}\label{e.exe.hamiltonian.anis}
H&=& 1/2 \sum_{\underline{i}} \left[ \left( \sum_{\underline{k}_1}
f_x S_{\underline{i}}^x S_{\underline{k}_1}^x
+ f_y S_{\underline{i}}^yS_{\underline{k}_1}^y + f_z
S_{\underline{i}}^zS_{\underline{k}1}^z\right) \right. \nonumber \\
& + & \left. \left( \sum_{\underline{k}_2} g_x S_{\underline{i}}^x S_{\underline{k}_2}^x
+ g_y S_{\underline{i}}^y S_{\underline{k}_2}^y + g_z
S_{\underline{i}}^z S_{\underline{k}_2}^z\right) - 2 h S_{\underline{i}}^z
\right] , \nonumber \\
\end{eqnarray}
where the index $\underline{k}_1$ runs over the first $Z_1^{(1)}$ nearest neighbors,
and the index $\underline{k}_2$ runs over the remaining $Z_1^{(2)}$ nearest neighbors
for each spin $S_{\underline{i}}$. According to the different values of $Z_1^{(1)}$ and
$Z_1^{(2)}$ this Hamiltonian describes various possible models. For instance, if $Z_1^{(1)} = 2$
and $Z_1^{(2)} = 1$ Eq. (\ref{e.exe.hamiltonian.anis}) describes $XYZ$ models on ladders
constituted by two coupled linear chains; the case $Z_1^{(1)} = 4$ and $Z_1^{(2)} = 1$
corresponds to models defined on two coupled square lattices. In both cases, every spin in
each chain (plane), besides the usual interactions within a single chain (plane) involving $Z_1^{(1)}$
nearest neighbors, is also coupled to $Z_1^{(2)}$ companion spins located on the second chain (plane).
The instance $Z_1^{(1)} = Z_1^{(2)} = 1$ corresponds to a $XYZ$-type model defined on a linear
chain with nearest-neighbor interactions of alternating strengths. The factorization properties of this latter model have been studied by G. L. Giorgi in the limit $f_z = g_z = 0$ that, belonging to the $XY$ symmetry class, can be solved exactly \cite{Giorgi}.

\begin{figure}[t]
\centering
\includegraphics[width=8.5cm]{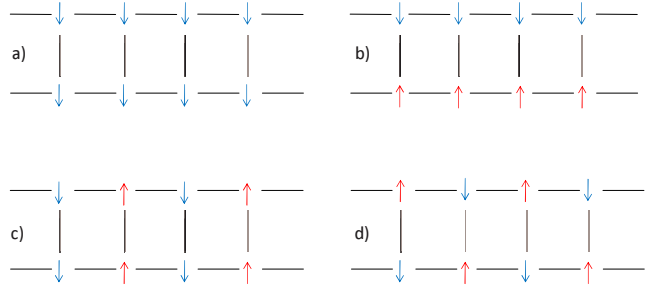}
\caption{(Color online) Summary of the possible magnetic arrangements along a single
direction ($x$ or $y$) characterizing the candidate factorized ground
states of models with spatially anisotropic interactions. The case considered
corresponds to a ladder of two coupled one-dimensional spin chains with
coordination numbers $Z_1^{(1)} = 2$ and $Z_1^{(2)} = 1$.
Topmost row: Parallel (a) and antiparallel (b) ferromagnetic order.
Bottommost row: Parallel (c) and antiparallel (d) antiferromagnetic order.}
\label{f.disegno}
\end{figure}

As we have seen in Sec.~\ref{sec.Method}, the first step of the
method is to single out the magnetic order that minimizes the
energy associated to the candidate factorized ground state, at fixed
magnetization along the $z$ direction (the direction of the external
field). Provided there are no effects of frustration, in the present
case we obtain that every possible hypothetical factorized ground state
can assume one of eight different magnetic orders. These eight
possibilities stem from the four possible orders of \eq{fourcases},
that the system can support when the effects of $g_\alpha$ can
be neglected, duplicated according to the two possible arrangements,
parallel or antiparallel, induced by the interactions associated to
$g_\alpha$. The four possible orderings along a single direction
are represented in Fig.~\ref{f.disegno} for a model of two transversally
coupled one-dimensional spin chains.

By imposing the minimization of the energy per site, \eq{e.energy}, the kind
of order actually present in the ground state can be determined comparing
the antiferromagnetic and ferromagnetic net interactions along the $x$ and $y$
axes, defined as
\begin{eqnarray}
\label{e.exe.netinter.anis}
\mathcal{J}^{A^\pm}_{\alpha} &=& - Z_1^{(1)} f_{\alpha} \pm Z_1^{(2)} g_{\alpha}\,; \nonumber \\
\mathcal{J}^{F^\pm}_{\alpha} &=& + Z_1^{(1)} f_{\alpha} \pm Z_1^{(2)} g_{\alpha} \, , \nonumber
\end{eqnarray}
where the $\pm$ sign discriminates between parallel/antiparallel ordering with respect to
$g_{\alpha}$. As a function of the net interactions, the type of
magnetic order present in the system is determined by the value of the minimum, i.e. of
$\mu=\min\{\mathcal{J}^{F^+}_x,\mathcal{J}^{F^-}_x,\mathcal{J}^{A^+}_x,\mathcal{J}^{A^-}_x,
\mathcal{J}^{F^+}_y,\mathcal{J}^{F^-}_y,\mathcal{J}^{A^+}_y,\mathcal{J}^{A^-}_y\}$.
Accordingly, we have the following correspondences:
\begin{equation}
\mu =\left\{ \begin{array}{ccl}
  \mathcal{J}^{F^+}_x & \Rightarrow & \hbox{Parallel ferromagnetic order along $x$ ;}  \\
  \mathcal{J}^{F^-}_x & \Rightarrow & \hbox{Antiparallel ferrom. order along $x$ ;}  \\
  \mathcal{J}^{F^+}_y & \Rightarrow & \hbox{Parallel ferrom. order along $y$ ;}  \\
  \mathcal{J}^{F^-}_y & \Rightarrow & \hbox{Antip. ferrom. order along $y$ ;}  \\
  \mathcal{J}^{A^+}_x & \Rightarrow & \hbox{Parallel antiferrom. order along $x$ ;}  \\
  \mathcal{J}^{A^-}_x & \Rightarrow & \hbox{Antip. antiferrom. order along $x$ ;}  \\
  \mathcal{J}^{A^+}_y & \Rightarrow & \hbox{Parallel antiferrom. order along $y$ ;}  \\
  \mathcal{J}^{A^-}_y & \Rightarrow & \hbox{Antip. antiferrom. order along $y$ .}
  \end{array}\right.  \label{e.exe.eightcases}
\end{equation}
Obviously, also in the presence of spatial anisotropies we may
observe that, in terms of the net interactions, the fact that in the
system there is no frustration (and especially no frustration
arising from the spatial anisotropies) implies that $\mu = - Z_1^{(1)}
|f_\alpha| - Z_1^{(2)}|g_\alpha|$, where $\alpha=x,y$ according to which
axis is characterized by the given magnetic order.

Without loss of generality, let us consider the situation in which
the system realizes a parallel antiferromagnetic order along the $x$ direction,
see Fig.~\ref{f.disegno}(c). All other cases can be treated analogously.
Following once again the steps described in Sec.~\ref{sec.Method}, and in
complete analogy with \eq{e.condition}, it is possible to derive the
set of conditions that the orientation $\theta$ of the E-SQUO must
satisfy simultaneously in order for the system to admit a factorized
energy eigenstate:
\begin{eqnarray}
\cos^2 \theta & = & \frac{f_y + f_z}{f_x + f_z} \label{e.exe.theta.anis.1}\\
\cos^2 \theta & = & \frac{g_y - g_z}{g_x - g_z} \label{e.exe.theta.anis.2}
\end{eqnarray}
For the two conditions to be satisfied simultaneously the two
r.h.s. must coincide and must assume values in the interval
$[0,1]$. Hence, by equating them it is possible to derive an equation
that plays the same role of Eqs. (\ref{e.exe.cantix}--\ref{e.exe.cferroy}).
In the hypothesis that the model under investigation satisfies both
Eqs.~(\ref{e.exe.theta.anis.1}--\ref{e.exe.theta.anis.2}), what is
left to prove is that the associated factorized state be indeed the ground
state. Following the route illustrated in Sec.~\ref{sec.Method} it is
not difficult to verify that the following inequalities must hold:
\begin{eqnarray}
(f_x + f_y) ( f_x + f_z) \ge 0 \; ;&& f_x  \ge - f_z \; ;
\label{e.exe.suffcondition.anis.1} \\
(g_x + g_y) (g_x - g_z) \ge 0 \; ;&& g_x   \le   g_z \; .
\label{e.exe.suffcondition.anis.2}
\end{eqnarray}
It is straightforward to establish that they are both satisfied, provided
there are no frustration effects. Namely, for spins interacting via
couplings of the first kind $f_\alpha$ (referring to Fig.~\ref{f.disegno},
these are the intra-chain interactions), the second of
Eqs. (\ref{e.exe.suffcondition.anis.1}) yields that the denominator
in the r.h.s. of \eq{e.exe.theta.anis.1} is non-negative and,
therefore, the corresponding numerator must be non-negative and not
exceeding the denominator. Hence, we obtain $f_x \ge f_y$. On the other hand,
from the first of Eqs. (\ref{e.exe.suffcondition.anis.1}) it follows
that  $f_x \ge - f_y$ which, in turn, implies $f_x \ge |f_y|$.
The latter relation is in agreement with the hypothesis of absence of
frustration. Furthermore, for pairs of neighboring spins interacting with
a coupling of the second kind $g_\alpha$ (referring to Fig.~\ref{f.disegno},
these are the inter-chain interactions), the second of Eqs. (\ref{e.exe.suffcondition.anis.2})
yields that the denominator in the r.h.s. of \eq{e.exe.theta.anis.2} must
be non-positive. Hence, also the corresponding numerator must be
non-positive and therefore $g_x \le g_y$. Simultaneously,
the first of Eqs. (\ref{e.exe.suffcondition.anis.2}) implies $g_x \le -
g_y$ and therefore $g_x \le |g_y|$, again in complete agreement with
the hypothesis of absence of frustration. Collecting all these
results we can conclude that in the absence of frustration, if a system
described by the model Hamiltonian
\eq{e.exe.hamiltonian.anis} satisfies simultaneously all conditions
in Eqs.(\ref{e.exe.theta.anis.1}--\ref{e.exe.theta.anis.2}), there
exist a factorizing field $h_f$ at which the ground state is fully
factorized and characterized by a parallel antiferromagnetic order
along $x$. The factorizing field is determined according to the procedure
described in Sec.~\ref{sec.Method}, and it is not difficult to verify
that it reads:
\begin{equation}
\label{e.exe.hf.anis}
h_f=\frac{1}{2}\sqrt{(\mathcal{J}_x^{A^+}-\mathcal{J}_z^{F^+})(\mathcal{J}_y^{A^+}-\mathcal{J}_z^{F^+})}
\; .
\end{equation}
The exact form of the factorized ground state reads
\begin{eqnarray}
\label{e.antiferroparrallelolungox}
\ket{\Psi_f}& = & \bigotimes_{\underline{k}} \ket{\psi_{{\underline{k}}}} \, ,\\
\ket{\psi_{{\underline{k}}}}&=&\left(
\cos(\theta/2)\ket{\uparrow^u_{\underline{k}}} +e^{i
\varphi_{\underline{k}} } \sin(\theta/2)
\ket{\downarrow^u_{\underline{k}}} \right) \nonumber \\
& \otimes & \left(
\cos(\theta/2)\ket{\uparrow^l_{\underline{k}}} +e^{i
\varphi_{\underline{k}} } \sin(\theta/2)
\ket{\downarrow^l_{\underline{k}}} \right) \, .
\end{eqnarray}
In \eq{e.antiferroparrallelolungox}, the superscript ``$u$'' (``$l$'')
denotes the upper (lower) chain of the ladder model considered
in Fig.~\ref{f.disegno}(c). We observe that
$\varphi_{\underline{k}}=|\underline{k}| \pi$ does not depend on
the choice of the chain, implying that the two generic $k$-th spins in
both legs of the ladder are in the same state equipped with an
antiferromagnetic order with respect to the nearest neighbor sites
along the corresponding chain. Analogous results hold, with the
appropriate trivial modifications, when considering the other possible
magnetic orders compatible with factorization (See \eq{e.exe.eightcases}).

\section{Conclusions and outlook}
\label{sec.Concl}

The present work has been motivated both by the need for
exact solutions to complex many-body quantum spin models,
in particular by the question ``when is a mean-field solution exact?'',
and by the necessity to acquire improved knowledge and control on the
structure of quantum correlations for potential technological
applications of spin systems. We have introduced a simple
and rather powerful all-analytic general method to determine the
conditions for the existence and the properties of fully factorized
(fully separable) ground states in translationally invariant quantum
spin models with general two-body exchange interactions and subject
to external fields.

The theory developed in the present paper builds on and extends
the scheme originally introduced in Ref. \cite{theory}.
It provides a readily useful procedure to establish, in terms
of the Hamiltonian parameters, whether the ground state is entangled
or completely factorized and, in the latter case, what is the
exact form of the state, the expression of the ground-state energy
and of the magnetic observables, and the type of ordered phase
(magnetic order) compatible with factorization. The task is achieved
by exploiting tools imported from quantum information theory such as
the formalism of single-qubit unitary operations introduced in
Refs. \cite{GiampaoloIlluminati,GiampaoloIlluminatiVerrucchi}
and the corresponding entanglement excitation energies, whose
property of vanishing if and only if a quantum ground state is
totally factorized plays a central role in the analysis of ground-state
separability. Using these tools we have completed the program initiated
in Ref. \cite{theory} by determining a general set of exact relations
and inequalities that, for frustration-free systems, allow to establish
rigorously the occurrence of all types of possible factorized ground states.

We have applied the analytic method to spin-$1/2$ models with general
Heisenberg-like exchange interactions of arbitrary range
(short, finite, long), either isotropic or anisotropic, defined on
regular lattices of arbitrary dimensions. The method is insensitive to
the size of the system and applies rigorously in the thermodynamic limit
as well as for finite lattices. The method allows to establish novel
exact, fully factorized ground state solutions of generally non exactly
solvable models, corresponding to nontrivial sets of values of the
Hamiltonian parameters. Key results include, for instance, the rather
straightforward determination of the existence and form of factorized ground states
in $XYZ$-type models with different types of short-range and long-range
interactions on cubic lattices, a task that would be of formidable complexity
if tackled with numerical techniques or resorting to the simple-minded direct
method based on the product ansatz.

The complete factorization diagram for nearest-neighbor Heisenberg-like
anisotropic models might be particularly useful for those technological
implementations which employ spin systems as information processors
exploiting their ground state entanglement \cite{Bose}. We have provided in
Fig.~\ref{f.factor} a ``minefield'' map of the unentangled
working points that need to be avoided, when engineering the couplings and tuning the
external fields, in order to achieve satisfactory quantum transmission performances.
On the other hand, there are also alternative tasks in quantum information which
instead require as initial working points factorized states, e.g. for engineering,
via suitable dynamics, specific classes of long-distance entangled states useful
for quantum state transfer and quantum teleportation \cite{AdolfBose} or the instantaneous
creation of strongly entangled graphs or cluster states involving a mesoscopic or
macroscopic number of spins for one-way quantum computation \cite{RaussendorfBriegel}:
In these instances the minefield may turn into a treasure map. The analysis of ground-state
factorization would be particularly useful in the study of quantum spin models on open-end
lattices. Such systems are not translationally invariant, and therefore the analytic
method requires to be suitably extended to this type of instances. This generalization is
under way and should be in reach in the near future.

Notably, according to a general theorem by Kurmann, Thomas, and M\"uller on factorization \cite{Kurmann},
given any Hamiltonian of the form \eq{e.Hamiltonian} involving higher spins $S > 1/2$,
the ground state of that spin-$S$ system is fully factorized at the same value of the external field $h=h_f$
\eq{e.factorizingfield}, at which factorization occurs in the corresponding spin-$1/2$ model. Therefore, the method and the results derived in the present paper have a much wider scope of application and can be straightforwardly generalized to interacting systems with arbitrarily high values of the spin, provided
they are of the same Hamiltonian structure as in the spin-$1/2$ case.

Absence of frustration effects, either of geometric or dynamical nature, has been essential to
the theoretical scheme derived in the present work. For instance, we have determined the
factorization diagram for specific examples of frustration-free systems endowed with isotropic
interaction terms of different spatial ranges, and for systems with nearest-neighbor anisotropic
interactions. However, it is clear that there are many extremely interesting instances of more complex
systems that would in general be subject to frustration effects. Continuing a step-by-step strategy
of applying the general method to models of increasing complexity, natural further stages of
investigation would include models subject to frustration, e.g., models on linear chains with competing isotropic interactions of different range such as a 1-D $XYZ$-type model with nearest-neighbor ferromagnetic interactions and next-to-nearest neighbor antiferromagnetic interactions. Increasing further the degree
of complexity, one could then consider models with competing anisotropic interactions of different ranges,
that according to the type of anisotropy could be subject to both geometrically- and dynamically-induced frustrations.

When considering frustrated models, except for very special cases, even when the considered system admits a fully factorized state as an energy eigenstate, the set of conditions derived in the present work are not sufficient to establish that such an eigenstate is indeed the ground state of the system. In fact, a competition arises between factorization, which requires a magnetic ordering, and frustration, which tends to destroy magnetic orders, ultimately leading to a chaotic correlating behavior. The latter unavoidably tends to enhance ground-state entanglement, thus suppressing the occurrence of situations where the most
favorable state, in terms of energy content, is completely uncorrelated. In such cases, a naive application of the direct method based on the factorized ansatz, for instance to models including anisotropic interaction terms of different magnetic types on many different spatial scales (or even on all scales), leads unavoidably to overestimate the factorizing effect of the balancing between interactions and external fields and thus to badly incorrect predictions on the occurrence of fully factorized ground states in these classes of models, as done, unfortunately, in the second part of Ref. \cite{Giorgi}. In a forthcoming work \cite{Inprep} we will present a systematic and thorough study of some important classes of frustrated quantum spin models in order to establish the conditions for the occurrence of true ground state factorization and to characterize it in terms of compatibility thresholds with frustration.

An intriguing issue that arises in connection with frustration is the relationship between factorization and the existence of ordered phases. For frustration-free Hamiltonian systems that conserve the even-odd parity, spontaneous symmetry breaking and the existence of an ordered phase are necessary conditions for ground state factorization, because factorized states have no definite parity. However, in the presence of frustration there may exist factorized energy eigenstates without ground state factorization. It would then be very interesting to understand whether in these situations there is still a causal relation between factorization and the existence of phase transitions in the system. Factorization might then be used as an heuristic tool to gain insight in the phase diagram of frustrated quantum systems.

It would also be important to understand whether chiral interactions can enhance factorization beyond the limits imposed by the presence of frustration effects. In fact, since factorization can be seen as ``mean field becoming exact'', it seems to require quite naturally a balancing between interactions and external fields as the only possibility for its occurrence. However, one cannot exclude, in principle, that other mechanisms might lead to the same effect without requiring the presence of external driving fields.

Future investigations will be concerned with the application of the analytic method to the determination of the spectrum of factorization, that is the study of the occurrence of factorized excited states in different classes of quantum spin Hamiltonians, and the implications of factorization diagrams on the structural and informational properties of quantum spin models for potential applications in quantum technology.

\acknowledgments

We acknowledge CNR-INFM Research and Development Center "Coherentia", INFN,
and ISI Foundation for financial support. One of us (F.I.), would like to
thank Ylenia D'Autilia for invaluable exchanges.

\bigskip

\begin{appendix}
\section*{Appendix: Determination of the magnetic orderings compatible with ground-state factorization}\label{appendixfourcases}

In Sec.~\ref{sec.Method} we have determined the general conditions
for ground state factorization by selecting as candidate factorized
ground states only those characterized by a ferromagnetic or
antiferromagnetic order along the $x$ or $y$ directions depending on
the value of the minimum value $\mu$ in the set of the net interactions
[see \eq{fourcases}], and neglecting all other possible magnetic arrangements.
In the following, we will prove that indeed, given any spin
Hamiltonian of the type \eq{e.Hamiltonian}, if a factorized
ground state exists, then it is characterized by one of the four
above-mentioned magnetic orders.

Let us consider the most general factorized state compatible with the
constraint of a site-independent magnetization $M_z$:
\begin{equation}
\label{e.candidate.gen} \ket{\Psi_f}= \bigotimes_{\underline{k}}
\ket{\psi_{{\underline{k}}} } \;, \; \; \; \;
\ket{\psi_{{\underline{k}}}}=\cos(\theta/2)
\ket{\uparrow_{\underline{k}}} +e^{i \varphi_{\underline{k}} }
\sin(\theta/2) \ket{\downarrow_{\underline{k}}} \; , \nonumber
\end{equation}
where the local spin state $\ket{\psi_{{\underline{k}}} }$ is
eigenstate of
\begin{equation}\label{e.oz.gen}
\bar{O}_{\underline{k}} = \cos \theta
S^z_{\underline{i}} + \cos \varphi_{\underline{k}} \sin\theta S^x_{\underline{i}}+
\sin \varphi_k \sin\theta S^x_{\underline{i}} \\
\end{equation}
with eigenvalue equal to $1/2$. Let us suppose that
there exists an Hamiltonian of the type \eq{e.Hamiltonian} that
admits, for some particular value of the external field $h=h_f$, the
state in \eq{e.candidate.gen} as an eigenstate. Under these hypotheses
we must have that the projection of the factorized eigenstate onto the
four-dimensional Hilbert space of a pair of spins in the lattice
must be an eigenstate of the pair-Hamiltonian term
$H_{{\underline{i}},{\underline{j}}}$ defined as in \eq{e.couple}.

Following the same strategy as in Sec.~\ref{sec.Method} let us
introduce for each site ${\underline{i}}$ a set of three auxiliary
mutually orthogonal spin operators among which one, that we name
$A^z_{\underline{i}}$, coincides with $\bar{O}_{\underline{i}}$ :
\begin{eqnarray}
\label{e.axayaz.gen}
 A^x_{\underline{i}} &=& \cos \theta \cos \varphi_{\underline{i}}
        S^x_{\underline{i}}+ \cos \theta \sin \varphi_{\underline{i}}
        S^y_{\underline{i}}-\sin \theta S^z_{\underline{i}}; \nonumber \\
        A^y_{\underline{i}} &=& -\sin \varphi_{\underline{i}} S^x_{\underline{i}}
        +\cos\varphi_{\underline{i}} S^y_{\underline{i}}; \\
        A^z_{\underline{i}} &=& \sin \theta \cos \varphi_{\underline{i}}
        S^x_{\underline{i}}+ \sin \theta \sin \varphi_{\underline{i}}
        S^y_{\underline{i}}+\cos \theta S^z_{\underline{i}} \; .
        \nonumber
\end{eqnarray}
By inverting Eqs.~(\ref{e.axayaz.gen}), we can conveniently re-express
the conventional single-spin operators as functions of the new set
of operators $\{A^\alpha_{\underline{i}}\}$:
\begin{eqnarray}\label{e.sxsysz.gen}
S_{\underline{i}}^x  &=& A_{\underline{i}}^x \cos \theta \cos
\varphi_{\underline{i}} -A_{\underline{i}}^y \sin \varphi_{\underline{i}}
+ A_{\underline{i}}^z \sin \theta \cos \varphi_{\underline{i}}; \nonumber \\
S_{\underline{i}}^y  &=& A_{\underline{i}}^x \cos \theta \sin
\varphi_{\underline{i}} +A_{\underline{i}}^y \cos \varphi_{\underline{i}}
+ A_{\underline{i}}^z \sin \theta \sin \varphi_{\underline{i}};\\
S_{\underline{i}}^z  &=&  - A_{\underline{i}}^x \sin \theta
+A_{\underline{i}}^z \cos  \theta \, .  \nonumber
\end{eqnarray}
Substituting \eq{e.sxsysz.gen} into the pair  Hamiltonian we
obtain the expression for $H_{\underline{i},\underline{j}}$ at factorization as a
function of the sets of operators $\{A^\alpha_{\underline{i}}\}$ and
$\{A^\alpha_{\underline{j}}\}$, which reads
\begin{widetext}
\begin{eqnarray}
\label{e.pairhij.gen} H_{\underline{i},\underline{j}}&=&
A_{\underline{i}}^z A_{\underline{j}}^z \left[  \cos ^2 \theta J_z^r
+ \sin ^2 \theta \left( J_x^r \cos \varphi_{\underline{i}} \cos
\varphi_{\underline{j}} + J_y^r \sin \varphi_{\underline{i}} \sin
\varphi_{\underline{j}}\right) \right] -2 h_f^r \sin \theta \left[
{A_{\underline{i}}^x \left( A_{\underline{j}}^z -\frac{1}{2}\right)
+\left( A_{\underline{i}}^z-\frac{1}{2}\right) A_{\underline{j}}^x }
\right] \nonumber \\
& & +A_{\underline{i}}^x A_{\underline{j}}^x \left( J_z^r-2 h_f^r
\cos \theta\right)+ A_{\underline{i}}^y A_{\underline{j}}^y \left(
J_x^r \sin\varphi_{\underline{i}} \sin \varphi_{\underline{j}} +
J_y^r \cos \varphi_{\underline{i}} \cos \varphi_{\underline{j}}
\right) - h_f^r \cos \theta \left( {A_{\underline{i}}^z +
A_{\underline{j}}^z } \right) \nonumber \\ & & +\left(
A_{\underline{i}}^z A_{\underline{j}}^y \cos \theta
+A_{\underline{i}}^x A_{\underline{j}}^y \sin \theta
\right)\left(-J_x^r \cos \varphi_{\underline{i}} \sin
\varphi_{\underline{j}}+ J_y^r \sin \varphi_{\underline{i}} \cos
\varphi_{\underline{j}} \right) \\ & & + \left( A_{\underline{i}}^y
A_{\underline{j}}^z \cos \theta +A_{\underline{i}}^y
A_{\underline{j}}^x \sin \theta \right)\left( -J_x^r \sin
\varphi_{\underline{i}} \cos \varphi_{\underline{j}}+ J_y^r \cos
\varphi_{\underline{i}} \sin \varphi_{\underline{j}} \right) \,
,\nonumber
\end{eqnarray}
\end{widetext}
where, as usual, $r$ is the distance between sites ${\underline{i}}$ and ${\underline{j}}$,
$J_\alpha^r$ are the spin-spin couplings at distance $r$ and along direction $\alpha$, and

$h_f^r$ obeys to the following relation
\begin{equation}\label{e.hfij.gen}
2 h_f^r = \cos\theta \left( J_z^r - \cos{\varphi_{\underline{i}}}
\cos{\varphi_{\underline{j}}} J_x^r - \sin{\varphi_{\underline{i}}}
\sin{\varphi_{\underline{j}}} J_y^r \right) \; .
\end{equation}

>From  \eq{e.pairhij.gen}, one has that to ensure that the
projection of the factorized state onto the four-dimensional Hilbert
space of spins $S_{\underline{i}}$ and $S_{\underline{j}}$ is an
eigenstate of the pair Hamiltonian, it is necessary to require that the terms
associated to the operators $A_{\underline{i}}^x A_{\underline{j}}^y $,
$A_{\underline{i}}^y A_{\underline{j}}^x $, $A_{\underline{i}}^z
A_{\underline{j}}^y$ and $A_{\underline{j}}^y A_{\underline{i}}^z$
vanish. This condition yields that
\begin{eqnarray}
J_x^r \cos \varphi_{\underline{i}} \sin \varphi_{\underline{j}}&=&
J_y^r \sin \varphi_{\underline{i}} \cos \varphi_{\underline{j}} \; ,
\label{e.condition2.1.gen}   \\
J_x^r \sin \varphi_{\underline{i}} \cos \varphi_{\underline{j}} &=&
J_y^r \cos \varphi_{\underline{i}} \sin \varphi_{\underline{j}}\; .
\label{e.condition2.2.gen}
\end{eqnarray}
Solving \eq{e.condition2.1.gen} and \eq{e.condition2.2.gen} we
obtain that if $|J_x^r|\neq|J_y^r|$, then the phases $\varphi$ obey
to the following relations:
\begin{equation}\label{e.sol.condition2.1.gen}
 \varphi_{\underline{i}}=0,\pi/2 \; \; \; \varphi_{\underline{j}}=
 \varphi_{\underline{i}}+n \pi \; ,
\end{equation}
with $n$ integer. It is immediate to verify that these relations are
consistent only with a ferromagnetic or with an antiferromagnetic order
along the $x$ or the $y$ direction. This concludes the proof.
\end{appendix}

\end{document}